\documentclass[fleqn,usenatbib]{mnras}
\usepackage{siunitx}
\usepackage{xcolor}
\usepackage{newtxtext,newtxmath}

\usepackage[T1]{fontenc}

\DeclareRobustCommand{\VAN}[3]{#2}
\let\VANthebibliography\thebibliography
\def\thebibliography{\DeclareRobustCommand{\VAN}[3]{##3}\VANthebibliography}

\usepackage{graphicx}	
\usepackage{amsmath}	
\usepackage{enumitem}

\usepackage{subcaption}

\usepackage[
    starfontserif 
    ]{starfont}

\DeclareSymbolFont{starfontsym}{OT1}{sts}{m}{n}
\DeclareMathSymbol{\mathSun}{\mathord}{starfontsym}{115}
\DeclareMathSymbol{\mathMercury}{\mathord}{starfontsym}{102}
\DeclareMathSymbol{\mathVenus}{\mathord}{starfontsym}{103}
\DeclareMathSymbol{\mathTerra}{\mathord}{starfontsym}{76}
\DeclareMathSymbol{\mathvarTerra}{\mathord}{starfontsym}{108}
\DeclareMathSymbol{\mathMoon}{\mathord}{starfontsym}{100}
\DeclareMathSymbol{\mathvarMoon}{\mathord}{starfontsym}{97}
\DeclareMathSymbol{\mathMars}{\mathord}{starfontsym}{104}
\DeclareMathSymbol{\mathJupiter}{\mathord}{starfontsym}{106}
\DeclareMathSymbol{\mathSaturn}{\mathord}{starfontsym}{83}
\DeclareMathSymbol{\mathUranus}{\mathord}{starfontsym}{70}
\DeclareMathSymbol{\mathvarUranus}{\mathord}{starfontsym}{65}
\DeclareMathSymbol{\mathNeptune}{\mathord}{starfontsym}{71}
\DeclareMathSymbol{\mathPluto}{\mathord}{starfontsym}{74}
\DeclareMathSymbol{\mathvarPluto}{\mathord}{starfontsym}{72}

\title[IPS-informed heliospheric modelling for MPTA]{Interplanetary scintillation-informed heliospheric modelling for the MeerKAT Pulsar Timing Array 4.5\,yr dataset}

\author[Mishra et al.]{
Saurav Mishra,$^{1,2,3}$\thanks{E-mail: smishra@swin.edu.au (SM)}
Daniel J. Reardon,$^{1,2}$
Andrew Zic,$^{3,2}$
Matthew Bailes,$^{1,2}$
John Morgan,$^{4}$
\newauthor 
Atharva D. Kulkarni,$^{1,2}$
Matthew T. Miles,$^{2,5}$
Ryan M. Shannon,$^{1,2}$
Caterina Tiburzi,$^{6,7}$
\newauthor
Mark Cheung,$^{3}$
Michael Kramer,$^{8,9}$
Ruoyao Ni$^{10}$
\\
$^{1}$Centre for Astrophysics and Supercomputing, Swinburne University
of Technology, P.O. Box 218, Hawthorn, VIC 3122, Australia\\
$^{2}$OzGrav: The ARC Centre of Excellence for Gravitational Wave Discovery, Hawthorn VIC 3122, Australia\\
$^{3}$ CSIRO, Space and Astronomy, PO Box 76, Epping, NSW 1710, Australia\\
$^{4}$ CSIRO Space and Astronomy, Bentley, 6102, WA, Australia\\
$^{5}$ Department of Physics and Astronomy, Vanderbilt University, 2301 Vanderbilt Place, Nashville, TN 37235, USA\\
$^{6}$ INAF - Osservatorio Astronomico di Cagliari, via della Scienza 5, 09047 Selargius (CA), Italy\\
$^{7}$ INAF-Osservatorio Astronomico di Brera, via Brera 28, Milan, Italy\\
$^{8}$ Max-Planck-Institut f\"ur Radioastronomie, Auf dem H\"ugel 69, D-53121 Bonn, Germany\\
$^{9}$ Jodrell Bank Centre for Astrophysics, University of Manchester, Alan-Turing Building, Oxford Street, Manchester M13 9PL, UK\\
$^{10}$ Imperial Colledge London}

\date{Accepted XXX. Received YYY; in original form ZZZ}

\pubyear{\the\year{}}

\begin{document}
\label{firstpage}
\pagerange{\pageref{firstpage}--\pageref{lastpage}}
\maketitle

\begin{abstract}
Heliospheric density variations impart delays on pulse times of arrivals from millisecond pulsars. Improper modelling of these variations may affect gravitational wave detection and characterisation by pulsar timing arrays (PTAs). Currently, PTAs typically employ a time-varying, spherically symmetric heliosphere model, which does not capture the full spatial and temporal complexity of the heliosphere. Instead, we investigate whether a three-dimensional, time-dependent model of the inner heliosphere from interplanetary scintillation (IPS) measurements -- the IPS-UCSD model -- can be employed to mitigate the solar wind in PTA analyses. We applied the IPS-UCSD model to the MeerKAT PTA 4.5‑year dataset to assess whether it could correct for heliospheric density variations, and the impact on GW sensitivity compared to a spherically-symmetric model. We find that the model does not accurately correct for heliosphere-induced timing distortions, leading to bias in recovered GW parameters. Using simulations, we show that the spherically symmetric heliosphere model also fails to fully capture heliospheric density variations like those in the IPS-UCSD model. However, if interstellar dispersion measure (DM) variations are also modelled, then the heliospheric model errors are partially absorbed by DM variations, reducing contamination of the GW signal. Therefore we find that a time-varying spherically symmetric model is sufficient to mitigate the effect of heliospheric time delays on recovered GW results at typical PTA radio frequencies, provided other signal components are also modelled. We propose that the most precisely timed pulsars may be used to improve data-driven heliospheric density models in the future.

\end{abstract}



\section{Introduction}

Pulsars are precise natural clocks that can act as Galactic-scale detectors of nanohertz-frequency gravitational waves (GWs). GWs are expected to induce delays in pulsar times of arrival (ToAs) on the order of hundreds of nanoseconds (ns) over a timescale of years. Sources of nanohertz GWs include inspiralling supermassive black hole binaries (SMBHBs) and potentially early-Universe sources such as inflation or cosmic strings \citep[e.g.][]{Burke_Spolaor_2019}. The ensemble of all inspiralling SMBHBs creates a stochastic gravitational wave background (GWB), which induces delays in pulsar ToAs \citep{1983ApJ...265L..39H}. By timing an array of pulsars over many years ($\gtrsim 5$\,yr) and searching for spatial correlation across various pulsar ToAs that has a particular form, known as the Hellings and Downs correlation \citep[HD,][]{1983ApJ...265L..39H}, the GWB signal can, in principle, be detected \citep{1979ApJ...234.1100D}. 

GW detection at nanohertz frequencies is performed using a pulsar timing array (PTA), which involves regular observations of millisecond pulsars (MSPs) distributed across the sky for multiple years \citep{1990ApJ...361..300F}. These PTA datasets not only enable the detection of GWs but also provide insights into properties of the interstellar medium (ISM) and heliosphere. Multiple PTA groups, such as the Chinese Pulsar Timing Array \citep[CPTA,][]{Xu_2023}, the European Pulsar Timing Array \citep[EPTA,][]{EPTA_2023}, the Indian Pulsar Timing Array \citep[InPTA,][]{Tarafdar_2022}, the MeerKAT Pulsar Timing Array \citep[MPTA,][]{Miles_2022}, the North American Nanohertz Observatory for Gravitational Waves \citep[NANOGrav,][]{Agazie_2023_DR}, and the Parkes Pulsar Timing Array \citep[PPTA,][]{zic2023parkespulsartimingarray}, collectively monitor approximately 130 MSPs. The PTA collaborations share data to form the International Pulsar Timing Array \citep[IPTA,][]{Perera_2019}.

Recently, the PTA collaborations have reported evidence for a GWB signal in their data with weak to strong evidence. However, the signal remains below the 5$\sigma$ detection threshold \citep{allen2023internationalpulsartimingarray}. Achieving this threshold requires accurate modelling of stochastic delays, spin irregularities, interstellar medium (ISM) turbulence, and heliospheric variations to avoid systematic errors. Inaccurate modelling of the aforementioned stochastic delays (noise processes) could lead to false detections of the GWB signal \citep[see, e.g.,][]{Tiburzi_2015, Goncharov_2021, 2024MNRAS.532.4026D} or bias the  recovered GW spectrum \citep[see, e.g.,][]{2022MNRAS.516..410Z, Reardon_2023_model, 2023ApJ...956...14D, 2025ApJ...990...85D}.

The plasma along the pulsar-Earth line-of-sight (LOS) induces radio-frequency–dependent dispersive delays ($\Delta t_{disper}$) in ToAs that are quantified by the dispersion measure (DM), defined as

\begin{equation}
    \centering
    \mathrm{DM} = \int_0^{L_a} n_{e}\, dl
\end{equation}
where $n_{e}$ is the free electron density along the LOS to the pulsar at distance $L_a$. The corresponding delay is
\begin{equation}
    \Delta t_{disper} = K_{D}^{-1}\frac{\mathrm{DM}}{\nu^{2}}
\end{equation}
where \textbf{$K_{D} = 2.41 \times 10^{-4}\ \mathrm{MHz}^{-2}\,\mathrm{pc}\, \mathrm{cm}^{-3}\,\mathrm{s}^{-1}$} is the dispersion constant and $\nu$ is the observing radio frequency \citep[see, e.g.,][]{lorimer_kramer, kulkarni2020dispersionmeasureconfusionconstants}.

The plasma in the pulsar-Earth LOS consists of plasma in the ISM \citep[see, e.g.,][]{2013MNRAS.429.2161K}, the heliosphere \citep[see, e.g.,][]{You_2007} -- the magnetised, supersonic plasma that completely fills the solar system, and around any binary companions, if relevant \citep[see, e.g.,][]{2019MNRAS.490..889P}. The changes in DM due to the ISM are related to the relative motion of the turbulent spatial structure in the ISM. The change due to such structures can be up to 10$^{-3}\, \mathrm{pc\,cm^{-3}}$ \citep[see, e.g.,][]{1993ApJ...404..636B, 2004MNRAS.353.1311H, 2013MNRAS.429.2161K}. These changes in DM are very well modelled by a temporal power spectrum with a power law \citep[see, e.g.,][]{2013MNRAS.429.2161K}. Additionally, for typical separations of pulsars, these variations can be assumed to be completely independent from one pulsar to the next.

The content of the heliosphere also causes significant dispersive delays for pulsars, especially for those with small ecliptic latitudes (ELAT) when they are observed close to the Sun \citep[see, e.g.,][]{You_2007, Tiburzi_2019, Susarla_2024}. The heliosphere is mostly filled with the charged stream of particles, which are coming out of the Sun. In PTA literature, these charged particles are referred to as Solar Wind (SW). However, in heliophysics, the term solar wind is often used to refer to the background outflow, while transient structures such as coronal mass ejections (CMEs) and co-rotating interacting regions are treated separately \citep[][]{Talpeanu_2022}. Therefore, heliospheric delay is the appropriate term for all delays caused by heliospheric plasma, whether background stream particles, or CMEs. In this paper, we use the term heliospheric delay, however, we use the notation ``SW'' for mathematical expressions of the noise model labels throughout this paper for consistency with recent work in the PTA literature.   

The dispersive delay due to the heliospheric plasma at typical L-band PTA observing radio frequencies ($\nu\sim$1400\,MHz) can be on the order of 1\,$\mu$s–100\,ns \citep[See, e.g.,][]{2007MNRAS.378..493Y, Tiburzi_2019, Tiburzi_2021, Kumar_2022}. Given the current sensitivity of PTAs, which can measure delays much smaller than 100\,ns \citep{Miles_2022}, it is crucial to accurately account for these heliospheric dispersive delays. To first order, the heliosphere is spherically symmetric, reducing in density with distance from the Sun as an inverse square law. This causes a very strong annual variation which peaks when the pulsar passes close to the Sun.

Traditionally, a static spherically-symmetric heliosphere model is been used to correct heliospheric delays in PTA data sets\footnote{Some PTA groups also use DMX to model chromatic delays. In DMX, a DM value is fit over a time bin spanning $\sim$ 1\,hour to weeks \citep{2017ApJ...841..125J}. DMX is expected to absorb the heliospheric dispersive delays, so, there is no need to model the heliosphere separately.}. This model assumes that the heliospheric electron density ($n_{e}$) varies inversely with the square of the radial distance ($r$), that is, $n_{e}(r) = n_{e}^{\mathrm{1\,AU}}(1\,\mathrm{AU}/r)^{2}$, where $n_{e}^{\mathrm{1\,AU}}$ (cm$^{-3}$) is the fixed electron density at 1\,AU \citep{Edwards_2006, Aksim_2019, 2019ApJ...872..150M}. The DM due to the heliosphere along the pulsar-Earth LOS at a certain angular separation $\theta$ (pulsar-observatory-Sun angle) is \citep{Edwards_2006}
\begin{equation}
    \mathrm{DM}_{SW} = n_{e}^{\mathrm{1\,AU}}\,(1\,\mathrm{AU})^{2}\,\frac{\pi-\theta}{\mathrm{r_{Earth}}\sin\theta},
    \label{DM_sph}
\end{equation}
where $\mathrm{r_{Earth}}$ is the distance from the Sun to the Earth. This simple model captures the first-order behaviour of a heliosphere-induced delay that peaks sharply at solar conjunction (the closer to the ecliptic plane the pulsar is, the sharper and greater the peak is). As radio telescope sensitivity has improved the precision of ToA measurements, it has become evident that this static model does not adequately account for heliosphere-induced delays \citep[see, e.g.,][]{2007MNRAS.378..493Y}. \citet{Tiburzi_2019, Tiburzi_2021, Hazboun_2022, niţu2024gaussianprocessesapproachfittingtimevariable, Susarla_2024} used a spherically symmetric heliosphere model with time-varying $n_{e}^{\mathrm{1\,AU}}$ (further detail in Section \ref{Stochastic noise}). 

Currently, stochastic variations of the density at 1\,AU are modelled using the Fourier basis\footnote{\citet{niţu2024gaussianprocessesapproachfittingtimevariable, hazboun2025nanograv125yeardataset} modelled stochastic variations of the density at 1\,AU using a time basis and covariance matrix}, typically assuming a power-law model of the power spectral density (PSD) of the variations \citep{Hazboun_2022}. This modelling of a stochastic heliospheric density variations is therefore very similar to the approach taken to account for density variations in the ISM. It should be noted, however, that this approach does not take advantage of the known structure of the heliosphere and its variation with time. Moreover, unlike the ISM, heliospheric variations experienced by one pulsar cannot be assumed to be independent of those experienced by others. Rather, the picture is of LOS from different pulsars at different times traversing a common heliosphere which is varying spatially and temporally in complex ways. This means that the heliospheric dispersive delays from observations of different pulsars is spatially correlated, meaning that it can introduce noise that is correlated across multiple pulsars. This spatial correlation can partially mimic a GWB, potentially leading to spurious GW detection \citep{Tiburzi_2015}.

The current model to account for the delays due to the heliosphere used by PTAs lacks the complexity of the real heliosphere structures \citep[see, e.g.,][]{1996Ap&SS.243...87C, 1998GeoRL..25....1M, 2020ApJS..246...36A, 2020SoPh..295..151P}. For instance, the overall structure of the heliospheric density varies over the (approximately 11-year) solar cycle, with approximate spherical symmetry at solar maximum changing to lower density over the poles at solar minimum \citep{1980Natur.286..239C}.  In addition, transients in heliosphere \citep[e.g.][]{zhang2020earthaffectingsolartransientsreview} such as CMEs and stream and co-rotating interaction regions add further time-varying complexity that are not included in the current heliosphere model (used in the PTA analysis). 

Work has been done to compare the spherically symmetric time varying heliosphere models with independent heliospheric density measurements in order to verify the accuracy of the models in PTA analyses. 
\citet{Hazboun_2022} compared the sampled values of $n_{e}^{\mathrm{1\,AU}}$ inferred in PTA analyses with Advanced Composition Explorer (ACE\footnote{ACE is a NASA explorer designed to study the heliosphere. \href{https://www.swpc.noaa.gov/products/ace-real-time-solar-wind}{https://www.swpc.noaa.gov/products/ace-real-time-solar-wind}}) measurements of $n_{e}^{\mathrm{1\,AU}}$. Similarly, \citet{Susarla_2024} compared $n_{e}^{\mathrm{1\,AU}}$ measurements inferred using stochastic modelling of heliosphere with OMNI data\footnote{OMNI provides hourly-averaged measurements of heliospheric magnetic field and density, running since 1963 \citep{2014cosp...40E2435P}}. However, all of these data come from very close to Earth (the Sun-Earth L1 Lagrange point) and therefore cannot capture the full 3D structure of the heliosphere. heliospheric contributions to pulsar DM are dominated by the point of the LOS closest to the Sun, which can lie at any point within the inner heliosphere. 

Both \citet{You_2007} and \citet{Kumar_2022} have compared pulsar data with two very different heliosphere models. \citet{You_2007, 2012MNRAS.422.1160Y} used a combination of the structure fast and slow solar wind to account for heliospheric delays, whereas \citet{Kumar_2022} utilised Wang-Sheeley-Arge–ENLIL, a large-scale heliospheric model \citep{2015SpWea..13..676P}. However both were ultimately derived from magnetic field observations of the Sun, which can then be empirically converted into densities. Both demonstrated an improvement over simpler models. However, magnetometry is most accurate on the portion of the solar surface directly facing the Earth, whereas the heliosphere that most affects pulsar measurements originates from the poles or the limb of the Sun.

The measurements of heliosphere made using Interplanetary Scintillation (IPS) which has been used for decades to track heliospheric densities \citep{2023FrASS..1059166X}, provides a promising avenue for providing information on the full three-dimensional (3D) time-varying heliosphere. Measurements of the scintillation level of tens of sources scattered across the sky have been made daily by purpose-built instruments \citep{2011RaSc...46.0F02T}. These measurements can then be used to construct a 3D, time-varying model of the heliosphere \citep{2020FrASS...7...76J}.
These reconstructions have been validated by comparison with L1 measurements of density and velocity, and have been shown to be highly competitive with other widely-used space weather models \citep{2015SpWea..13..316J}. A particular advantage of IPS-based heliosphere models for pulsar timing applications is that the data from which these observations are derived are LOS integrated measurements made over multiple solar elongations and latitudes, and therefore directly probe the heliosphere that is most influential on pulsar observations.

IPS-based heliosphere reconstructions were recently compared with DM measurements of some pulsars observed very close to the Sun using LOFAR radio telescopes \citep{Tiburzi_2023}. These reconstructions were generated using the University of California San Diego (UCSD) heliosphere 3D tomography reconstruction technique. Compared to models derived from single-point \textit{in-situ} measurements close to the Earth, 3D tomography of the heliosphere from IPS can provide a more accurate model by incorporating more detailed information on the heliospheric origin ``transients'' such as stream interaction regions and coronal mass ejections throughout the heliosphere, while also providing a better representation of the overall 3D structure of the heliosphere. 

The \citet{Tiburzi_2023} study compared the heliosphere model derived from IPS-UCSD 3D tomography using pulsar observations at low radio frequencies (100-190 MHz) on a small data set, which motivated a study that uses 3D heliosphere reconstructions from IPS-UCSD with a much larger dataset from PTAs. However, other PTAs, including the MPTA, often observe pulsars at higher frequencies where the effect of delays due to the heliospheric variation is reduced. This motivates the study in this paper, which utilises IPS-UCSD 3D heliosphere reconstructions with a much larger observational dataset spanning radio frequencies from 0.8-1.7\,GHz.\\

This work is focused on taking advantage of the 3D time-dependent heliosphere model in PTA dataset where
\begin{enumerate}[leftmargin=*]
    \item We test whether this model can simply be assumed to be correct (mitigating the need to infer the true heliosphere from the PTA datasets). We utilise the MPTA dataset as it is one of the most sensitive PTA across the globe.  
    \item We utilise the 3D time-dependent heliosphere model to search for GWB on MPTA dataset and compare the recovered GWB parameters against those obtained using a spherically symmetric heliosphere model.
    \item We treat these heliosphere reconstructions as a ``plausible'' realisation of the heliospheric delays, apply it to simulations of PTA mock data, and see if existing heliosphere model in PTA are able to model it (sufficiently that GWs can be recovered without error.)
\end{enumerate}

In Section \ref{observation-methods}, we describe the dataset and methods used for modelling stochastic noise processes, simulating the heliospheric delays, and recovering spatial-correlations in the data like the GWB. In Section \ref{results} we describe our results including a comparison of the strength of the recovered GWB under different heliosphere models. We also assess how different heliosphere models affect the recovered spatial correlation of the GWB signal. We discuss and conclude the results in Section \ref{discussion}.

\section{Observations and Methodology}
\label{observation-methods}

\subsection{Stochastic noise modelling in Pulsar Timing Array}
\label{Stochastic noise}

Radio waves from pulsars experience multiple 
delays, including frequency-dependent delays caused by plasma in the ISM and heliosphere, Rømer delays, and Shapiro delays 
\citep{1990ApJ...361..300F, Edwards_2006, Hobbs_2006}. All of these physical phenomena must be accounted for when predicting the ToA of pulses through a model known as the timing model. These physical phenomena can be parameterised through timing model parameters, which thereby provide information about the astrophysical properties of pulsars, such as mass and rotation. The timing model fitting is performed using software packages such as \textsc{tempo2} \citep{Edwards_2006, Hobbs_2006} or PINT \citep{Luo_2021}, and is typically performed using frequentist least-squares methods.

However, the deterministic timing model is not a perfectly accurate description of ToAs; there can still be some correlated structure remaining in the difference between observed and predicted ToA, known as the timing residual ($\delta t$). The structure in the residuals may arise due to some unmodelled time-correlated stochastic signal ($\delta t_\mathrm{stoch}$), or any missing deterministic signal ($\delta t_{\mathrm{deter}}$) in the timing model along with some time-uncorrelated stochastic signal due to uncertainty in the measurement of the ToAs. Therefore, the residuals can be written as
\begin{equation}
    \delta t = \delta t_{\mathrm{stoch}} + \delta t_{\mathrm{deter}}  
\end{equation}

One origin of time-correlated stochastic signals in pulsar timing data is the turbulent plasma within the ISM and the heliosphere, which impart dispersive delays and scattering upon the pulsar radiation along the LOS. The GWB itself also induces stochastic variations in the ToAs \citep{1979ApJ...234.1100D}. These time-correlated stochastic signals typically have a red power spectrum.

In PTA data analysis, time-correlated stochastic signals are usually modelled as the Fourier basis using Gaussian process (GP) as 
\begin{equation} 
    \delta t_{\mathrm{stoch}} (t_i) = \sum_{l=1}^{N_{\mathrm{modes}}} \left[c_{l}\sin(2\pi f_{l}t_{i}) + d_{l}\cos(2\pi f_{l}t_{i}) \right] \times X_{\mathrm{stoch}}(\Theta_{t}),
\end{equation} 
\noindent where $N_{\mathrm{modes}}$ is the total number of Fourier frequencies being modelled for stochastic signals \citep{Lentati_2013}. $f_{l} = l/T_{\mathrm{span}}$ represents the $l$th frequency component, whereas $T_{\mathrm{span}}$ is the span of the dataset. $X_{\mathrm{stoch}}({\Theta_{t}})$ is an additional function that parametrises the dependence of the delays on other parameters such as the radio frequencies or the solar elongation. $c_{l}$ and $d_{l}$ are the Fourier coefficients. These coefficients are derived from a Gaussian distribution with variance defined using a PSD. The PSD can either be written as a ``free spectrum'', where the quadrature sum of the components $\sqrt{c_l^2 +d_l^2}$ is indepently fit for each frequency component, or constrained using a model such as a power law ($\mathrm{P_{stoch}}$)

\begin{equation}
    \label{PSD}
    P_{\mathrm{stoch}}(A, \gamma; f) = \frac{A^{2}}{12\pi^{3} f_{\mathrm{yr}}^{3}}\left(\frac{f}{f_{\mathrm{yr}}}\right)^{-\gamma} 
\end{equation}
\noindent where $A$ and $\gamma$ refer to the amplitude and the spectral index of the PSD function that describes the stochastic process. Here, the $c_l$ and $d_l$ Fourier coefficients are subject to the constraint that $\sqrt{c_l^2 +d_l^2}$ follows the power spectral density specified by Eq. \ref{PSD}, while the phase $\arctan(c_l/d_l)$ is left unconstrained and is marginalised out of the fit -- see \citet[e.g.][]{taylor2021nanohertzgravitationalwaveastronomer} for more details.

For each stochastic process, a $P_{\mathrm{stoch}}$ is defined using a PSD for that process. It is important to define PSDs for different stochastic processes differently to avoid mismodelling of various stochastic processes, and this is where $X_{\mathrm{stoch}}$ plays a key role. $X_{\mathrm{stoch}}$ is unity for achromatic time-correlated processes such as pulsar spin irregularities ($X_{\mathrm{RN}}$) or GWB, whereas for chromatic time-correlated processes (depending on observing radio frequency), $X_{\mathrm{stoch}}$ is defined in different ways, including:
\begin{subequations}
\begin{align}
\label{RN}
X_{\mathrm{RN}} &= 1 \\
\label{DM}
X_{\mathrm{DM}}(\nu) &= \left(\frac{\nu}{\nu_{\mathrm{ref}}}\right)^{-2} \\
\label{chrom}
X_{\mathrm{chrom}}(\nu,\beta) &= \left(\frac{\nu}{\nu_{\mathrm{ref}}}\right)^{-\beta}\\
\label{XSW}
X_{\mathrm{SW}}(\nu; \theta) &= \mathrm{DM_{SW}}(\mathrm{n_{e}^{1\,AU}} = 1; \theta)\left(\frac{\nu}{\nu_{\mathrm{ref}}}\right)^{-2},
\end{align}
\end{subequations}
\noindent where $\nu_{\mathrm{ref}}$ is the reference frequency, which is typically defined as 1400\,MHz. For this work, we are using the power-law description of various stochastic processes as discussed in the \citet{Miles_2024_model}. 
$X_{\mathrm{DM}}$ represents the change in variation in DM due to the turbulent nature of ISM \citep{1991ApJ...382L..27P} which is modelled as a stochastic time-correlated signal. $X_{\mathrm{chrom}}$ accounts for radio-frequency-dependent delays that may be due to small-scale structure present in ISM that causes multi-path scattering due to diffraction of the pulsar signal. This multi-path scattering is also modelled as a time-correlated stochastic variation, but with different frequency dependence $\mathrm{\beta}$. This may be set to a value of 4 if assuming standard cold-plasma theory \citep[e.g.][]{lorimer_kramer}, or be allowed to vary if relaxing the cold-plasma assumptions. $X_{\mathrm{SW}}$ represents changes in the dispersive heliospheric delays due to changes in $n_{e}^{\mathrm{1\,AU}}$ over time. Here, $\mathrm{DM_{SW}}$ is given in eq. \ref{DM_sph}, and accounts for the additional column density arising from the change in the pulsar-Sun elongation angle within a spherically-symmetric heliosphere.

There are also ``deterministic'' signals ($\delta t_{\mathrm{deter}}$) such as chromatic Gaussian jumps/dips, which arise from deviations from power-law stochastic processes in the ISM due to the presence of small-scale structures, annual variations in chromatic signals caused by Earth's orbital motion around the Sun, exponential-like delays due to sudden pulse profile shape changes, and mean delays due to the heliospheric variations \citep{Reardon_2023_model}. {We use the term ``deterministic'' here following standard parlance in the PTA noise modelling literature, but note that these terms are not physically deterministic. Instead these ``deterministic'' terms denote processes that can be written parametrically in the time domain, rather than described statistically through some description of its power spectrum.}A detailed description of all these deterministic signals modelled on the MPTA dataset can be found in \citet{Miles_2024_model}. For this work, the mean delay due to the heliospheric variations, is particularly relevant as we fit for the mean value of $n_{e}^{\mathrm{1\,AU}}$ using the dispersion measure from the variations in the heliosphere based on eq.~\ref{DM_sph}.

We also model time-uncorrelated stochastic signals. These stochastic signals behave as excess white noise, as the uncertainty on any ToA derived from standard pulsar timing procedures arises largely from the thermal noise of the telescope. However, there are additional sources of white noise that must be accounted for. The white noise term consists of EFAC, EQUAD, and ECORR. EFAC and EQAUD model the unaccounted systematics and imperfections in the pulse template fitting, such as those due to the observing and backend system of the radio telescope, and ECORR accounts for short timescale pulse shape variations known as pulse jitter \citep[see, e.g.,][]{2014MNRAS.443.1463S, Parthasarathy_2021, kulkarni2024insightchromaticbehaviourjitter}. The white noise is modelled by modifying the uncertainty of ToA measurements as done in see e.g., \citet{Reardon_2023_model}.  For the MPTA DR2 \citep{Miles_2024_model} we only have one receiver and backend combination across the data set. We explain this now, in the context of other PTA analyses where the white noise parameters are defined for each combination. 

The GWB at nanohertz frequencies induces a stochastic delay in all pulsars. These delays are common to all pulsars and are spatially correlated according to HD correlation relation. The HD correlation is defined using a two-point correlation function, the ORF, $\Gamma_{ab}$. The correlation due to GWB is written as \citep{1983ApJ...265L..39H}:

\begin{equation}
    \Gamma^{\mathrm{HD}}_{ab} = \frac{1}{2}(1+\delta_{ab}) - \frac{1 - \cos\zeta_{ab}}{4}\left[ \frac{1}{2} - 3\ln\left( \frac{1 - \cos\zeta_{ab}}{2}\right) \right] 
\end{equation}
\noindent where $\zeta_{ab}$ is the sky-separation angle between pulsar $a$ and pulsar $b$. However, for modelling the GWB, we need to add information about spatial correlation due to the GWB to the PSD function. Therefore, the power-law PSD due to the GWB is written as:
\begin{equation}
    P_{\mathrm{HD}} = P_{\mathrm{stoch}}(A_{\mathrm{GWB}},\gamma_{\mathrm{GWB}})\times \Gamma^{\mathrm{HD}}_{ab}.
\end{equation}

Once the parameters describing various processes are defined, the model parameters are inferred using Bayesian inference with \textsc{enterprise} \citep{2019ascl.soft12015E} and \textsc{enterprise$\_$extensions} \citep{enterprise} using \textsc{ptmcmcsampler} \citep{van_Haasteren_2009, van_Haasteren_2012, van_Haasteren_2014, justin_ellis_2017_1037579} as described in \citet{Miles_2024_model}. We define prior function over the modelled parameters describing various stochastic and determinstic signal. Typically, the timing model parameters are analytically marginalised out, leaving only the posterior probability distributions of the parameters
for the various noise processes to be determined. The most important parameters for this work are $A_{\mathrm{GWB}}$ and $\gamma_{\mathrm{GWB}}$. However, using P$_{\mathrm{HD}}$ requires a large amount of computational resources and time. So for a sensible rough estimate of $A_{\mathrm{GWB}}$ and $\gamma_{\mathrm{GWB}}$, the cross-correlation is ignored, that is the $A_{\mathrm{GWB}}$ and $\gamma_{\mathrm{GWB}}$ are inferred as a uncorrelated common red noise (CRN) using the pulsar auto-correlations only. We call the fitted CRN parameters $A_{\mathrm{CRN}}$ and $\gamma_{\mathrm{CRN}}$ \citep{Reardon_2023, Miles_2024_model, Gersbach_2025}.

After obtaining the values of $A_{\mathrm{CRN}}$ and $\gamma_{\mathrm{CRN}}$, a frequentist detection statistic is employed, known as the optimal statistic (OS), to fit for the value of $A_{\mathrm{GWB}}$ using the determined inter-pulsar covariances/correlations, as described in \citet{Chamberlin_2015, Gersbach_2025}. The OS can also be used to search for another correlations due to common-spectrum signals with e.g., monopolar or dipolar patterns \citep{2018MNRAS.481.5501C, 2020MNRAS.491.5951H}. A monopole correlation can exist, for example, because of an error in the observatory clock which is used as a reference for ToAs measurement, whereas a dipole correlation can be, for example, due to errors in the position and motion of planets. The \textsc{enterprise$\_$extensions} contains the OS code which has been used in this work and the same code also calculates the inter-pulsar covariances/correlations.

The first heliosphere model considered in this paper is the spherically symmetric heliospheric model. This heliosphere model consists of two components: a deterministic component SW$_{\mathrm{mean}}$ that is modelled using Eq.\ref{DM_sph}, where $n_{e}^{1\,AU}$ is fitted as a free parameter, and the stochastic component that uses a GP on $n_{e}^{1\,AU}$, SW$_{\mathrm{GP}}$. The PSD describing variations in $n_{e}^{1\,AU}$ is assumed to follow a power-law as given in Eq. \ref{PSD}, with an additional function defined as $X_{\mathrm{SW}}$ as described in Eq.~\ref{XSW}. In Section \ref{IPS-SWM}, we describe the application of the IPS-UCSD 3D reconstructed heliospheric model to the PTA noise model framework described above, as an alternative to the spherically symmetric model.

\subsection{MeerKAT PTA dataset and noise models}
\begin{figure*}
        \centering 
        \includegraphics[width=\textwidth]{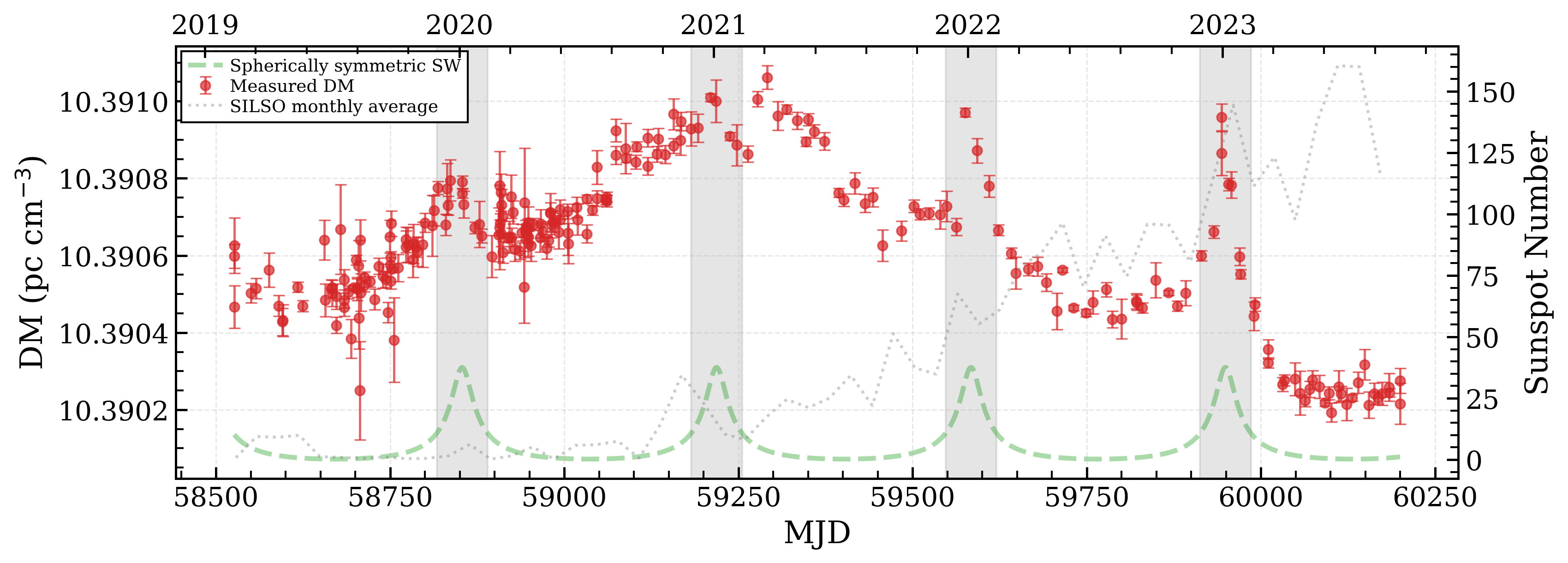}
    \caption{
    DM measurements from PSR~J1909$-$3744 using MeerKAT L-band observations. The grey regions show when the pulsar is within 40 degrees of the Sun. There are annual variations in the DM, e.g. clear DM enhancements around MJDs 58850, 59200, 59600, and 59950, which coincide with epochs of increased solar activity and small Sun-pulsar angular separations. At these times the change in the DM are of the order of 10$^{-4}\,\mathrm{pc\,cm^{-3}}$, which is the expected change in DM due to the heliospheric plasma. 
    The green curve represents a static spherically symmetric heliosphere model with $\mathrm{n_{e}^{1\,AU}}$ = 4.909 $\mathrm{cm^{-3}}$, offset by 10.39\,$\mathrm{pc\,cm^{-3}}$ to bring these variations to a similar level as the DM measurements shown in red. The  amplitude of these annual DM variations appears to increase through the dataset. This is due to the increase in solar activity, indicated by the Sunspot number\protect\footnotemark in grey on alternate y axis.  
    The long term variations in DM are associated with the changes in the interstellar plasma. Note that in the timing model used to measure the DMs, no model is included to account for heliospheric dispersive delays.}
    \label{fig:J1909-DM}
\end{figure*}
\footnotetext{The sunspot numbers are counted through the Sunspot Index and Long-term Solar Observations \href{https://www.sidc.be/SILSO/home}{https://www.sidc.be/SILSO/home}}

In this work, we use the 4.5$-$year dataset released as MPTA DR2 \citep{Miles_2024_model} spanning from 2019-2023. 83 MSPs are part of MPTA DR2, and for each pulsar, the ToAs are collected using the MeerKAT radio telescope \citep{2016mks..confE...1J} 
in South Africa, observing in the L-band radio frequency range (856-1712\,MHz). The details of the MPTA observing strategy can be found in \citet{Miles_2022}.  

MPTA DR2 has been searched for a uncorrelated CRN process (e.g., a GWB) in the data \citep{Miles_2024}. For this work, we begin with a similar search for noise processes, following the analysis methods from MPTA DR2. We perform searches for a spatially-uncorrelated CRN using following 3 heliosphere models: 
\begin{enumerate}[label=\alph*), leftmargin=*]
    \item ``SW$_{\mathrm{mean}}$+SW$_{\mathrm{GP}}$'' based on the MPTA DR2, 
    \item ``SW$_{\mathrm{IPS}}$'' for each pulsar,
    \item ``SW$_{\mathrm{IPS}}$+SW$_{\mathrm{GP}}$'' for all pulsars.
\end{enumerate}
where ``SW$_{\mathrm{mean}}$+SW$_{\mathrm{GP}}$'' is based on the spherically symmetric model described in Section \ref{Stochastic noise}. ``SW$_{\mathrm{IPS}}$'' represents the heliosphere model based on IPS-derived heliospheric variations using UCSD 3D-tomography as described in Section \ref{IPS-SWM}. ``SW$_{\mathrm{IPS}}$+SW$_{\mathrm{GP}}$'' represents a combined model of spherically symmetric heliosphere and IPS-UCSD derived heliospheric variations. 

After each CRN search, we reconstruct the time domain realisations \citep[see, e.g.,][Appendix A for the mathematical detail about time-domain realisations]{iraci2024pulsartimingmethodsevaluating} of various signals in the data to compare the inferred properties of the various modelled signal in presence of different heliosphere models. In brief, this reconstruction of time-domain signals is based on sampling the Fourier coefficients from the distributions defined by the inferred PSD parameters, conditioned on the information provided by the timing residuals. We use PSR\,J1909$-$3744 to compare the heliospheric delays, DM variations, and CRN signal because this pulsar is the most precisely timed pulsar in MPTA DR2, with median ToA error of 0.214$\mu s$. PSR\,J1909$-$3744 clearly exhibits heliospheric variations in the DM time series (with DM measured independently with \textsc{tempo2} at each observing epoch) as shown in Fig. \ref{fig:J1909-DM} due to its proximity to the ecliptic plane (ELAT=-15.15$^{\circ}$). Once each CRN search is complete, we also calculate the OS to investigate the strength of correlations using 3 ORFs: HD, dipole, and monopole correlations.

\subsection{Interplanetary scintillation and the heliosphere modelling using the UCSD 3D tomography}
\label{IPS-SWM}
\begin{figure}
        \centering 
        \includegraphics[width=\linewidth]{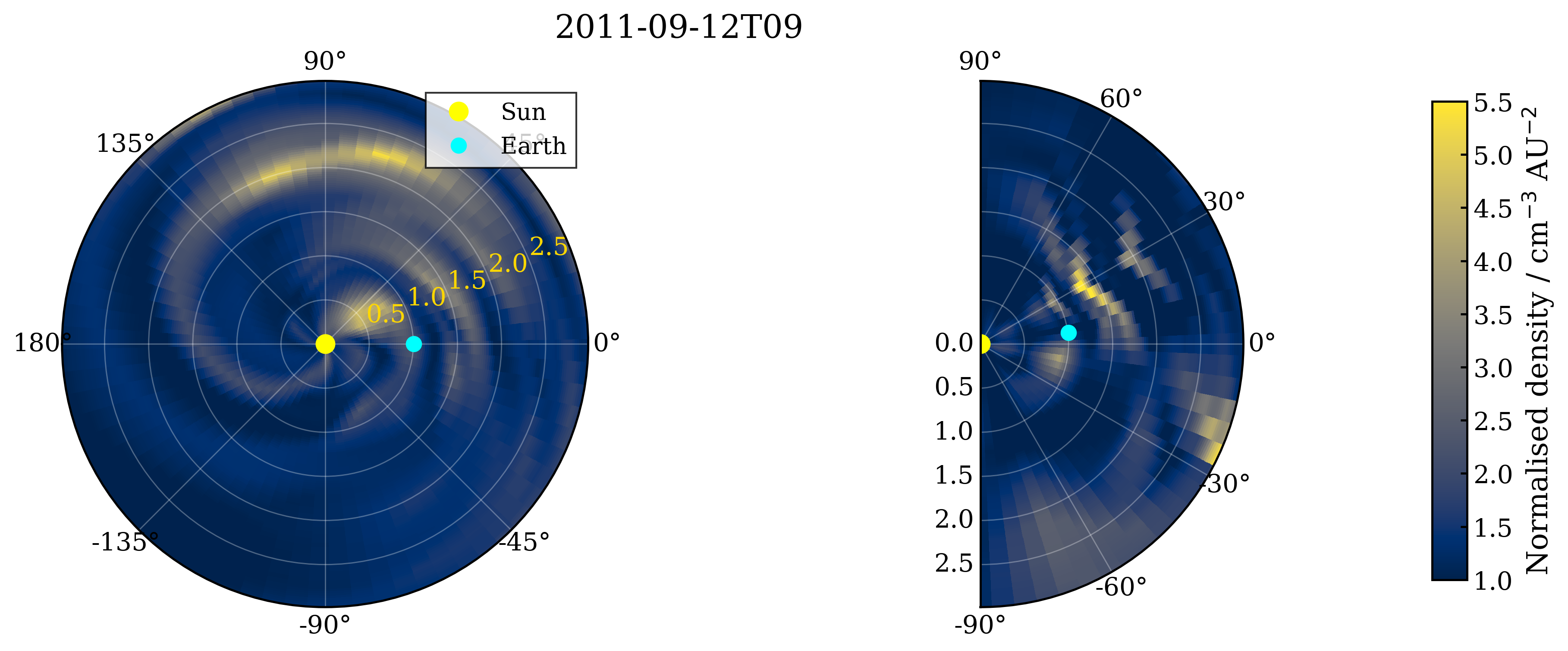}
       \includegraphics[width=\linewidth]{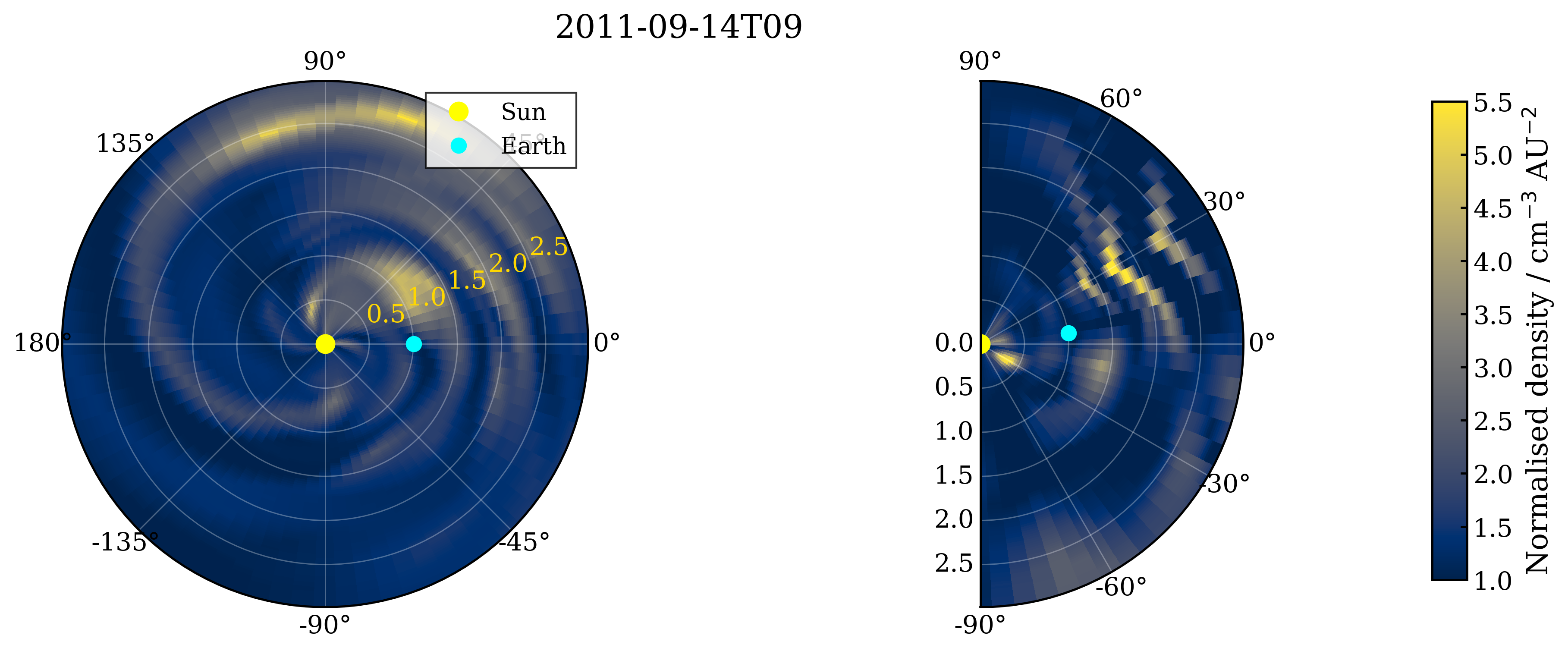}
    \caption{Two different cuts through volumetric density reconstructions at two different epochs. The upper plots and lower plots are separated by 2\,days.  The left plots show approximately ecliptic plane cut of the heliosphere (thr ecliptic pole and the SUn's rotational pole differe by 7$^{\circ}$ \citet{2007CeMDA..98..155S}) with longitude values in degrees, the right plots represent the perpendicular cut of the heliosphere with latitude values in degree. The inner boundary at 15$R_\odot$ is just visible, and the Earth is shown as a cyan dot. The colour bar represents the normalised electron density. Both the plots are shown in the Stonyhurst coordinate system (a coordinate system used in heliophysics, defined by the rotation of the Sun, with Sun at the origin. The z$-$axis or longitude +90$^{\circ}$ representing Sun's north pole, and the x$-$axis or longitude 0 and latitude 0 aligns with the projection of the Sun$-$Earth separation onto the Sun's equatorial plane \citep{sunpy_community2020}}).
    \label{fig:IPS-cut}
\end{figure}

Interplanetary Scintillation (IPS) has been used in studies of compact radio sources since its discovery \citep{1968Natur.217..709H}. IPS occurs when radio waves from a distant compact source scatter off electron density irregularities in the heliospheric plasma. This scattering produces a diffraction pattern at the observer, and the motion of these irregularities causes the observed intensity to fluctuate temporally \citep{1964Natur.203.1214H}. The strength of the IPS is quantified by the scintillation index -- the ratio of measured intensity fluctuations to the mean intensity of the compact source. 

The strength of IPS has been shown to be strongly correlated with the heliospheric density \citep[see Section 3 in][]{1993SoPh..148..153M} because both the scintillation index and the heliospheric density fluctuations depend strongly on the heliospheric distance (distance to plasma from the Sun), while the mean heliospheric density also depends strongly on the heliospheric distance \citep[see eq. 2 in][]{1993SoPh..148..153M}.
In addition, the scintillation enhancement factor -- defined as the ratio of measured scintillation index to the expected value in the weak scattering limit -- provides additional information. \citet{1986P&SS...34...93T} suggested a power-law relation between the scintillation enhancement factor and electron density at a given heliocentric distance. Therefore, measuring scintillation at low frequencies enables determination of electron density in the heliosphere.

The UCSD 3D tomography reconstruction technique can be used to map the density and velocity of the heliosphere at each point in space. This technique assumes an initial heliospheric velocity and density at radius $15\,R_{\odot}$ and evolves these values using a model that conserves mass and mass flux for a radially outward-moving heliospheric plasma, up to a radial distance of 3\,AU. The scintillation enhancement factor along the IPS source LOS is predicted, and an iterative least-squares fitting technique is applied to match the model-derived scintillation enhancement factors with measurements. After several iterations, a 3D map of heliospheric density and velocity is obtained. The reader is referred to the recent review of \citet{2020FrASS...7...76J} for further details.

For this work, we use these IPS-UCSD 3D tomography\footnote{\href{https://ips.ucsd.edu/}{https://ips.ucsd.edu/}} reconstructions of heliosphere density. These reconstructions are based on input data consisting of IPS velocities and densities obtained using a network of dedicated radio telescopes  Japan \citep{2011RaSc...46.0F02T} operating at 327 MHz. These input measurements then drive the UCSD 3D tomography code to produce 3D maps of the heliosphere. with a 3-hour cadence. Fig.~\ref{fig:IPS-cut} is an example cut of the model through the ecliptic plane, which shows clear, logarithmic (Parker) spiral structures. Since the Sun is rotating and the IPS-UCSD model assumes kinematic motion, these spirals will inevitably emerge as persistent features. There is also a transient feature of higher density visible in the ecliptic cut at a longitude of approximately 45$^\circ$. It is this combination of a persistent background heliosphere and transient features of higher density which makes IPS-UCSD measurements a realistic model of the heliosphere.

We employ the IPS-UCSD 3D heliosphere reconstruction in our effort to model delays due to the heliosphere variations in the MPTA DR2 dataset as follows. Heliospheric densities are determined along the LOS based on the pulsar's observing epoch by using the IPS-UCSD 3D tomography closest to that epoch (almost always within $\pm$3\,hours). The densities along the pulsar-Earth LOS are then integrated to calculate the total column density due to heliosphere (hereby called $\mathrm{DM_{IPS}}$), for every pulsar observation. A code has been developed to perform the integration of densities to estimate the DM due to the heliosphere using the IPS-UCSD 3D tomography model\footnote{The code will be shared upon request made to the corresponding author}.

The IPS-UCSD 3D tomography measurements contain outliers that are affected by radio frequency interference (RFI) or coronal mass ejections. Radio frequency interference can lead to higher modulation indices because of its time-variable behaviour, which in turn affects the modelled electron density. Coronal mass ejections, on the other hand, are unlikely to be modelled well enough to be predictive for our pulsar observations because they are spatially compact and short-lived.
\begin{figure}
   \centering        
        \includegraphics[width=\linewidth]{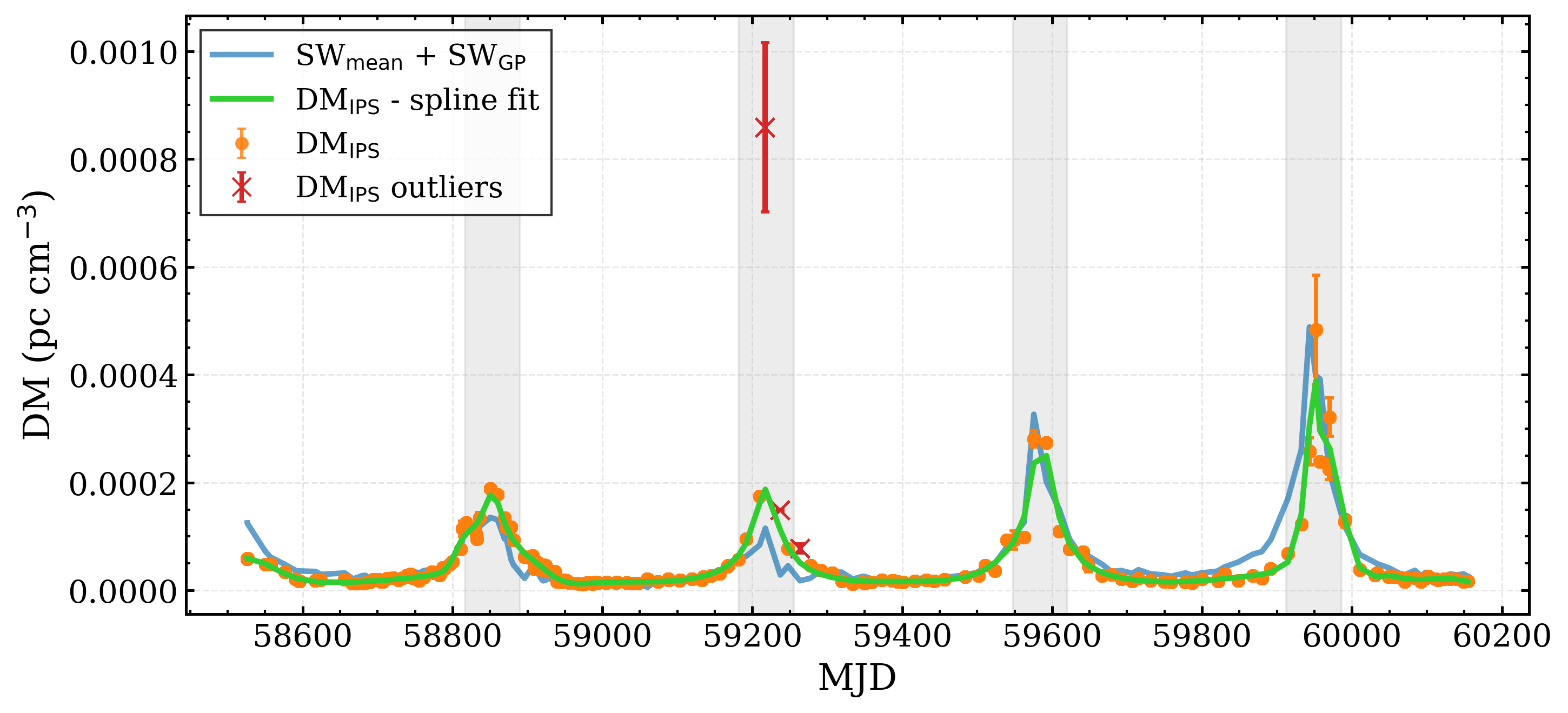}
        \label{fig:DM_in}
    \caption{Predicted heliospheric DM using the IPS-UCSD 3D tomography model for the PSR\,J1909$-$3744; Removal of outliers from IPS data and replacement of outliers by interpolating the good IPS data points in the density domain using spline interpolation. The x-axis shows MJD, the left y-axis shows DM. Orange and red points represent good and bad IPS-derived heliospheric delays data points, respectively. The blue line represents the DM derived from the spherically symmetric heliosphere model, while the green line represents the interpolated DM values used in this work. The gray strips represent when the pulsar is within the 40$^{\circ}$ of the Sun, leading to an increased heliospheric DM contribution.}
    \label{fig:interpolation}
\end{figure}

The outlying data points were identified by examining the error bars at each epoch, where we define the error bar as the standard deviation of density variations on the same pulsar-Earth LOS within $\pm$12\,hours. We exclude an epoch if its error bar exceeds 15 times the standard deviation of all error bars, and if $\mathrm{DM_{IPS}}$ exceeds 3 times the prediction from the PTA data using the ``SW$_{\mathrm{mean}}$ + SW$_{\mathrm{GP}}$'' model. These thresholds were chosen empirically to retain the bulk of the $\mathrm{DM_{IPS}}$ values while removing extreme outliers. To replace excluded epochs we perform a spline interpolation of the electron density values for the retained epochs (computing $\mathrm{n_{e}^{1\,AU}}$ using eq. \ref{DM_sph}). We apply the spline interpolation scheme to the $\mathrm{n_{e}^{1\,AU}}$ time series so that it preserves the long-term temporal structure of the retained data.

The $\mathrm{DM_{IPS}}$ measured from the IPS-UCSD 3D heliosphere reconstructions accounts for contributions up to 3\,AU. However, the relative contribution to heliosphere DM beyond 3\,AU can be significant for the LOS at larger solar elongations. To avoid spurious annual variations arising from this missing heliospheric contributions (beyond 3\,AU), we add a model for the DM beyond 3\,AU along the pulsar-Earth LOS. This extra DM from beyond 3\,AU is computed using a combination of the $\mathrm{DM_{IPS}}$ and a spherically symmetric heliosphere model \citep[similar to the approach taken by][]{Kumar_2022}. Concretely, we keep the IPS-UCSD 3D heliosphere reconstruction inside 3\,AU and model the DM from 3\,AU to the pulsar with a time-varying spherically symmetric heliosphere that uses $\mathrm{n_{e}^{1\,AU}}$ derived from the IPS-UCSD 3D heliosphere reconstructions. The procedure comprises two steps:
\begin{enumerate}[leftmargin=*]
    \item Estimate a time-varying spherically symmetric heliosphere from the maximum likelihood realisation of ``SW$_{\mathrm{mean}}$ + SW$_{\mathrm{GP}}$'' from the MPTA DR2 noise analysis. Compute the excess DM from 3\,AU to the pulsar as the difference between this spherically symmetric model and the DM obtained by integrating that same model only to 3\,AU. Add that excess DM to the $\mathrm{DM_{IPS}}$ to form an updated $\mathrm{DM_{IPS}}$, known as $\mathrm{DM_{IPS,updated}}$.
    \item Step 1 produces a composite heliosphere model combining the IPS‑UCSD 3D reconstruction with the spherically symmetric heliosphere inferred from the MPTA DR2 maximum‑likelihood ``SW$_{\mathrm{mean}}$ + SW$_{\mathrm{GP}}$'' realisation. Here we want an overall heliosphere that retains the IPS‑UCSD 3D reconstructed heliospheric densities but enforces a spherically symmetric for heliosphere DM beyond 3\,AU to pulsar. So, from the updated $\mathrm{DM_{IPS,updated}}$, compute the time-varying $\mathrm{n_{e}^{1\,AU}}$ using eq. \ref{DM_sph}. Smooth this $\mathrm{n_{e}^{1\,AU}}$ with a spline model to capture only the long-term variation, then use the smoothed $\mathrm{n_{e}^{1\,AU}}$ to calculate the excess DM from 3\,AU to the pulsar, relative to the spherically symmetric model. Add this smoothed excess DM to the original $\mathrm{DM_{IPS}}$, known as $\mathrm{DM_{IPS,final}}$.  
\end{enumerate}

In summary, we use a time-varying spherically symmetric heliosphere model to add any DM due to heliosphere beyond 3\,AU, but we base $\mathrm{n_{e}^{1\,AU}}$ for time-varying spherically symmetric heliosphere model on the IPS-UCSD 3D heliosphere reconstructions. This two-step procedure reduces the excess power at the $\mathrm{1\,yr}^{-1}$ frequency. We denote this $\mathrm{DM_{IPS,final}}$ as ``SW$_{\mathrm{IPS}}$''.

We also implement a hybrid model that uses ``SW$_{\mathrm{IPS}}$'' together with the ``SW$_{\mathrm{GP}}$'' component to absorb any remaining unmodelled heliospheric delays signal, which may be detectable in the pulsar data. This model is referred as ``SW$_{\mathrm{IPS}}$ + SW$_{\mathrm{GP}}$''.


\subsection{Simulation of mock Pulsar Timing Array datasets}
\label{simulation-section}
In this work, we also tested how the heliospheric dispersive delays may affect the measured spatial-correlations among pulsar pairs through simulated timing data sets. For simulations, we use the ``SW$_{\mathrm{IPS}}$'' as a realistic representation of heliospheric delays. We used the \textsc{libstempo} \citep{2020ascl.soft02017V} package, a python wrapper for \textsc{tempo2}, which includes routines for simulation of noise processes in ToAs to generate mock PTA datasets.

We produced 3 sets of simulations to test different aspects of the effect of heliosphere delays modelling on PTAs:

\begin{enumerate}[leftmargin=*]
\item \textbf{Simulation 1:} The first simulation involves generating ToAs with no stochastic red noise signals and no white noise, using the \textsc{idealtoas} function in \textsc{libstempo}, exactly matching the original MPTA DR2  MJDs and observed radio frequencies. We then inject heliospheric delays derived from the IPS-UCSD model, finally adding Gaussian white noise with a standard deviation of 100\,ns with EFAC set to 1. 
\item \textbf{Simulation 2:} We take the ToAs with IPS-UCSD 3D heliosphere reconstructions from \textbf{Simulation 1} and inject a GWB signal using the \textsc{creategwb} function in \textsc{libstempo} as described in \citet{Chamberlin_2015}. The GWB signal is injected as a common red noise signal as predicted by \citet{phinney2001practicaltheoremgravitationalwave}. The injected GWB signal has a PSD with $\log_{10}A=-14.7$ and $\gamma=13/3$ \citep[values reported in][]{Reardon_2023}, to be consistent with the current estimated level of the GWB from recent searches. We also inject DM variations as a stochastic red-spectrum GP which is associated to the stochastic interstellar DM variations. This DM variation is assumed to have a power-law PSD as described in Eq.\,\ref{PSD} with an additional function in Eq.\,\ref{DM}. This stochastic interstellar DM variation is denoted as DM$_\mathrm{GP}$ with the spectral index ($\mathrm{\gamma_{DM_{GP}}}$) fixed at $8/3$, as expected for a Kolmogorov turbulence spectrum \citep[see, e.g.,][]{1990ApJ...364..123F, lorimer_kramer}. The amplitudes of the power-law describing DM$_\mathrm{GP}$ are randomly drawn from a normal distribution with mean and  standard deviation derived from the sample of the interstellar DM variations power-law amplitudes reported in the MPTA DR2.
\item \textbf{Simulation 3:} The third simulation creates a PTA containing all pulsars in MPTA DR2 with ecliptic latitude, $|\mathrm{ELAT}|<16^\circ$ and a 13-years time baseline\footnote{The facility for IPS observations was upgraded in 2010 with a newly developed UHF radio telescope for IPS observations \citep{2011RaSc...46.0F02T}; therefore we use the timespan 2010--2023 for the simulation.} using multi-frequency observations with the same frequency range as the original MPTA DR2. heliospheric delays are injected based on the IPS-UCSD 3D reconstructed model with white-noise level set to 100\,ns, and setting EFAC=1.
\end{enumerate}

Note that in all of the above 3 simulations, we fix all the pulsar binary parameters except the period of the binary for the pulsars using the binary model. This was done to prevent divergences in the timing model due to non-linear fitting effects introduced by the simulated signals. In addition, to ensure the reproducibility of the simulated dataset, we simulated signals such as DM variations, white noise, and GWB using unique seed values. The aim of \textbf{simulations 1} and \textbf{2} are to characterise systematic effects when the heliospheric delays are not properly modelled and to show how this affects the recovered CRN parameters and spatial correlation of common signals using the optimal statistic. \textbf{Simulation 3} focuses on understanding the PSD of the stochastic components of the heliospheric delay: we use the IPS-UCSD 3D tomography reconstructed heliosphere as a realistic example of the heliospheric variations and recover SW$_\mathrm{GP}$ with a free-spectrum model to examine how the power is constrained at each Fourier frequency.

To model heliospheric delays in the above cases, we use \textsc{enterprise} routines and adopt the SW$_{\mathrm{mean}}$ + SW$_{\mathrm{GP}}$ model. We assume Fourier frequencies 1 / T$_{\mathrm{span}}$, 2 / T$_{\mathrm{span}}$, 3 / T$_{\mathrm{span}}$, \dots 120 / T$_{\mathrm{span}}$ (same number of components used in the simulation and the MPTA DR2) for the modelling, whereas the default \textsc{enterprise} heliospheric delays model uses frequency components equally spaced in logarithm space. To model the interstellar DM variations we use the DM$_{\mathrm{GP}}$ model, with eq. \ref{PSD} as the PSD definition and the additional function given in Eq. \ref{DM}. The Fourier-frequency grid for DM$_{\mathrm{GP}}$ is the same as for SW$_{\mathrm{GP}}$. We also consider a DM$_{\mathrm{free}}$ case, where a free spectrum is used to fit the noise at Fourier frequencies 1 / T$_{\mathrm{yr}}$, 2 / T$_{\mathrm{yr}}$,\dots, 10 / T$_{\mathrm{yr}}$  for modelling. For the GWB, the Fourier frequenices are chosen 1 / T$_{\mathrm{span}}$, 2 / T$_{\mathrm{span}}$, 3 / T$_{\mathrm{span}}$, \dots 30 / T$_{\mathrm{span}}$.

\begin{figure}
        \centering 
        \includegraphics[width=0.7\linewidth]{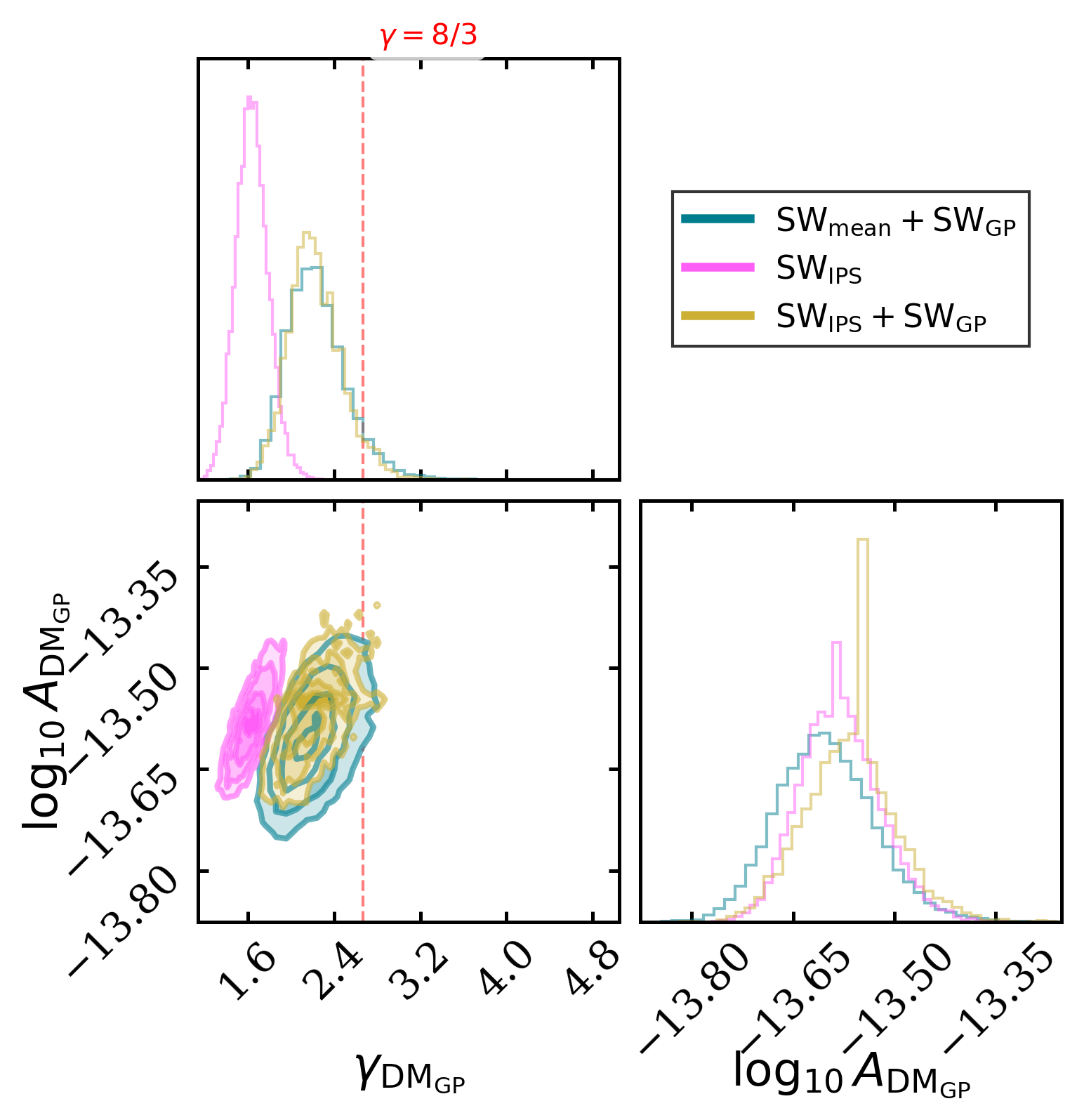}
        \includegraphics[width=0.7\linewidth]{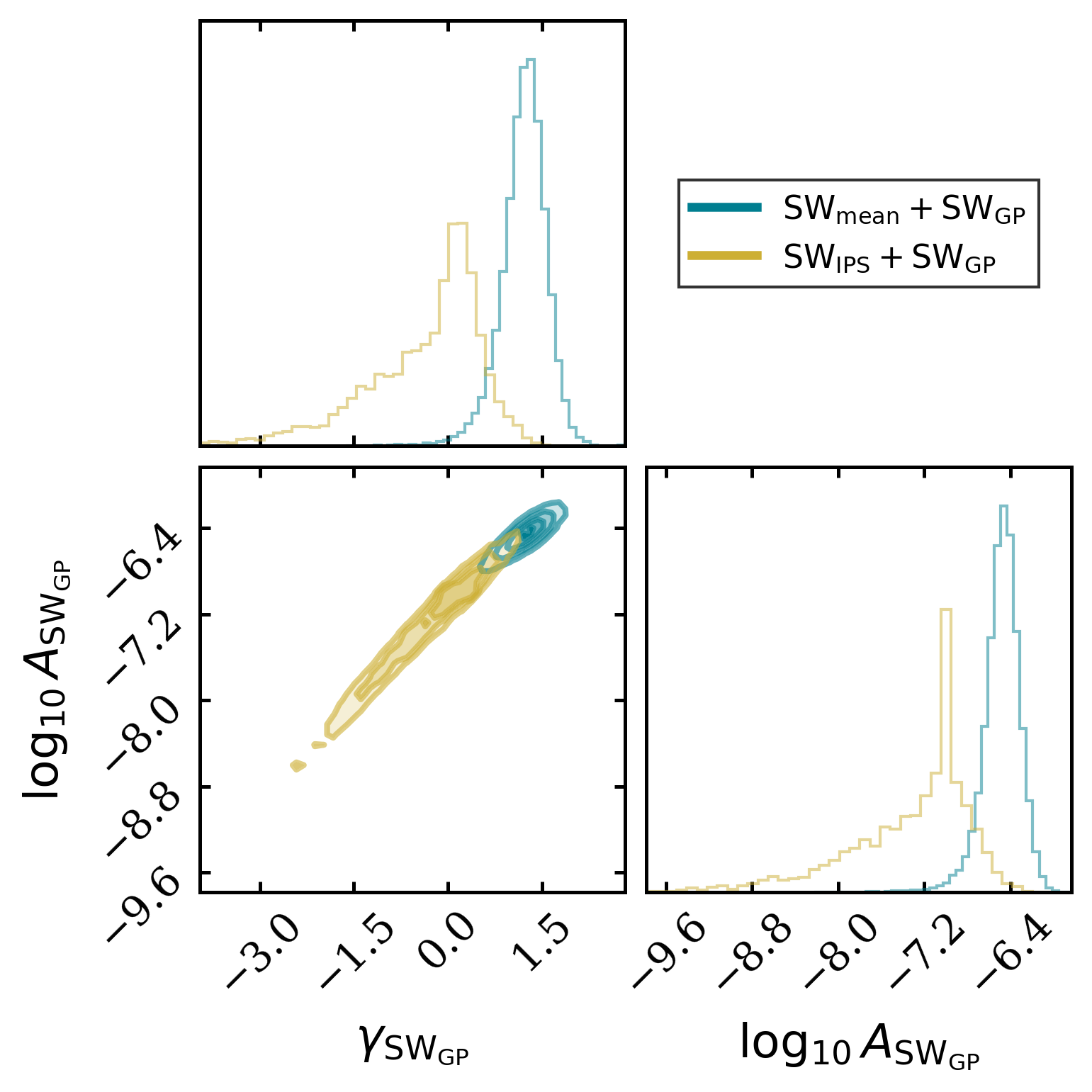}
        \centering
       \includegraphics[width=0.7\linewidth]{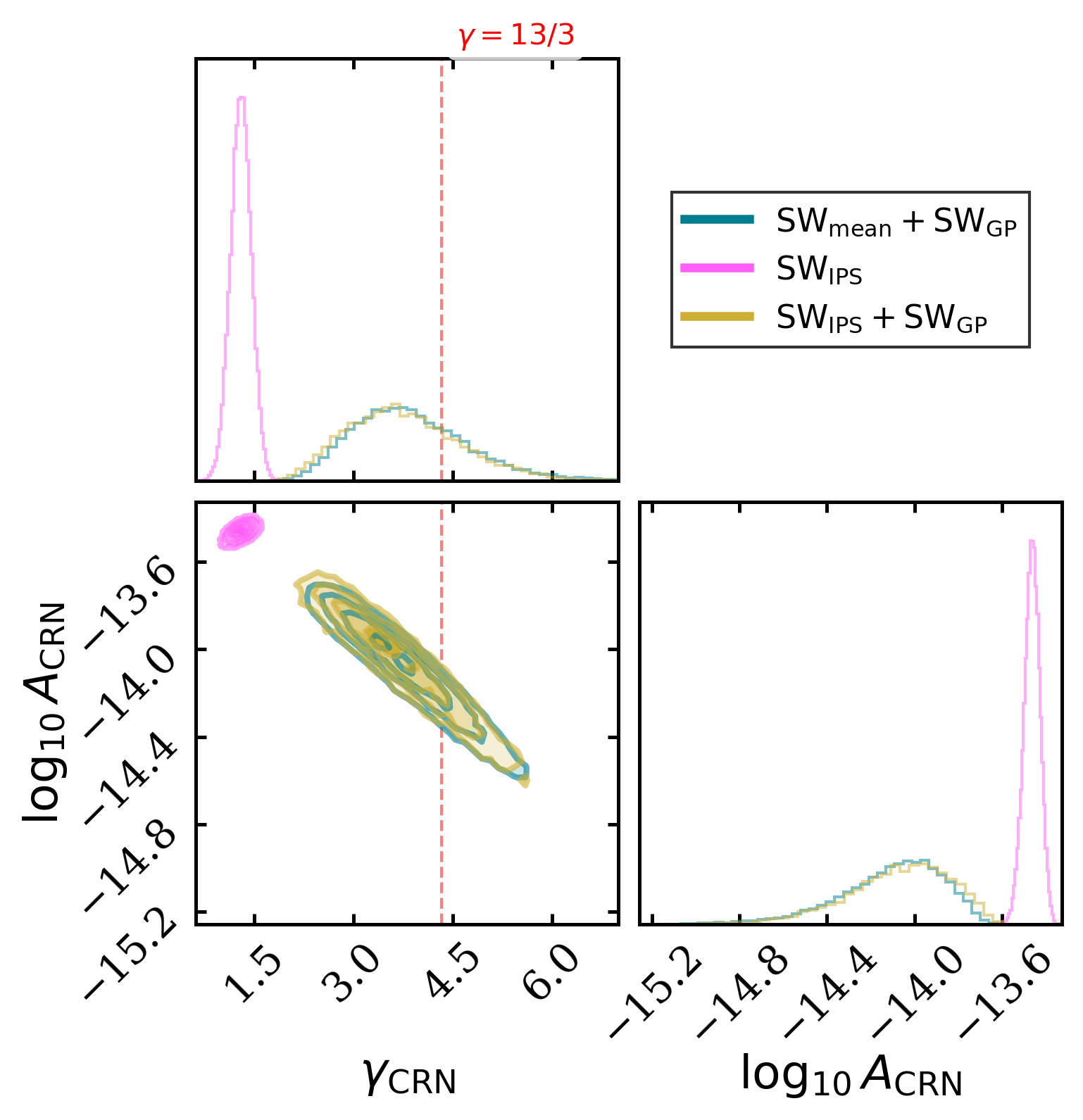}
    \caption{Posterior distributions of the inferred parameters (A and $\gamma$) of the power-law PSD describing a stochastic process. Top: the stochastic interstellar DM variations for the PSR~J1909$-$3744; Middle: the heliospheric density variations (i.e., $\mathrm{n_{e}^{1\,AU}}$) for the PSR~J1909$-$3744, bottom: the uncorrelated common red noise for the MPTA DR2. The teal, pink, and gold represent the case where ``SW$_{\mathrm{mean}}$+SW$_{\mathrm{GP}}$'', ``SW$_{\mathrm{IPS}}$'', and ``SW$_{\mathrm{IPS}}$+SW$_{\mathrm{GP}}$'' heliosphere models are used respectively. $\gamma$ = 8/3 corresponds to the Kolmogorov spectrum due to turbulence in the interstellar medium where $\gamma$ = 13/3 corresponds to gravitational wave background coming from a population of circular supermassive black hole binaries.}
    \label{fig:noise_param}
\end{figure}

\begin{figure*}
  \centering
  \begin{subfigure}{0.49\textwidth}
   \caption{Mean heliospheric DM + stochastic heliospheric DM}
    \includegraphics[width=\linewidth]{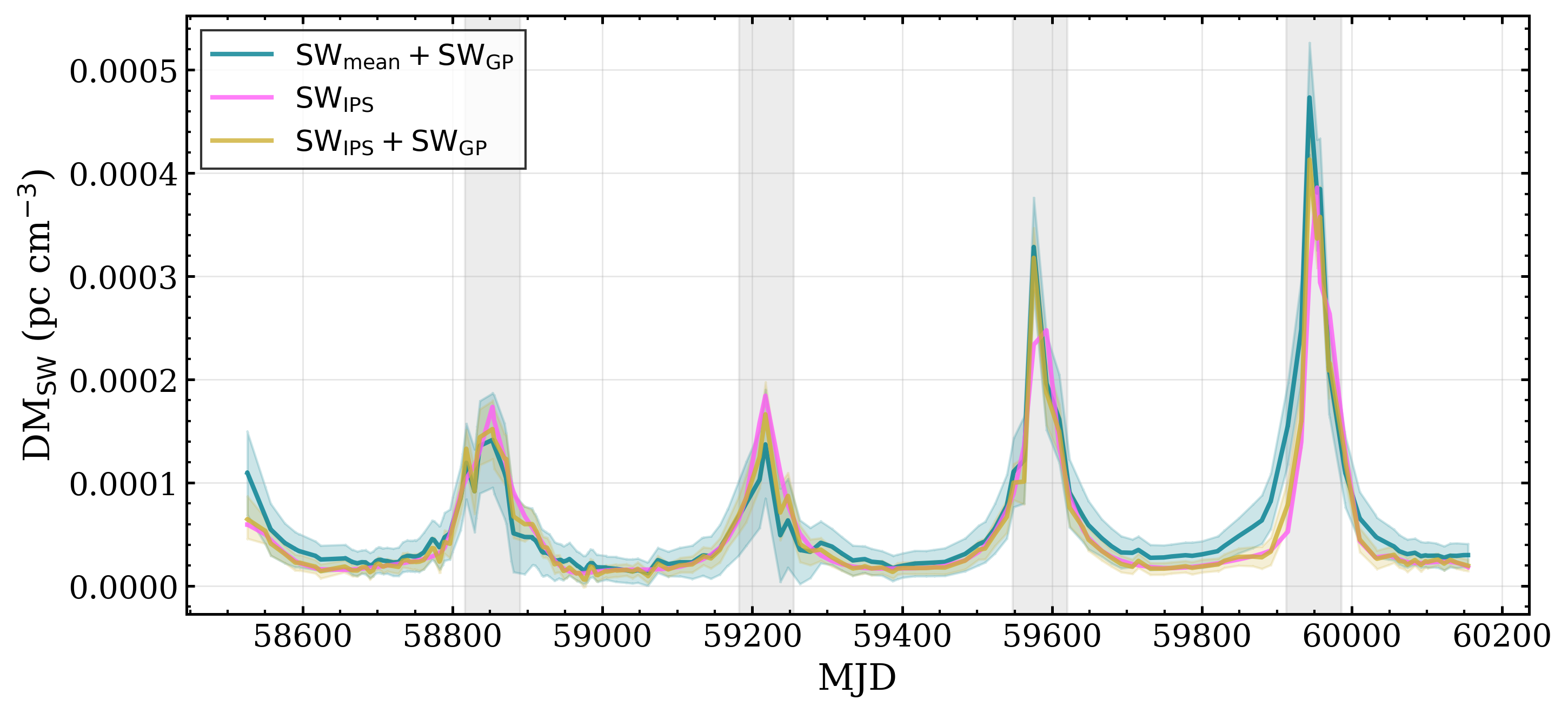}
    
  \end{subfigure}
  \begin{subfigure}{0.49\textwidth}
    \caption{Stochastic interstellar DM}
    \includegraphics[width=\linewidth]{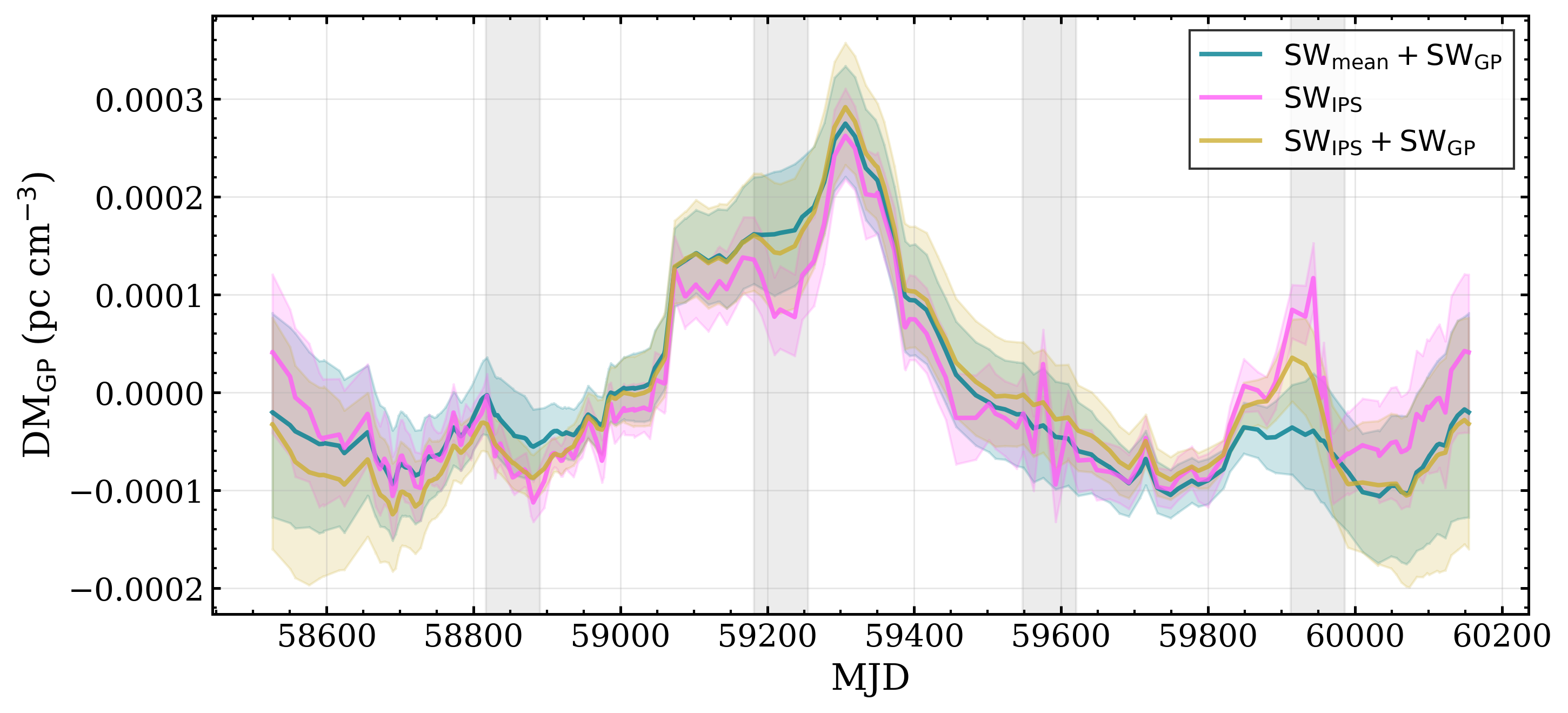}
    
  \end{subfigure}

  \begin{subfigure}{0.49\textwidth}
    \caption{Stochastic heliospheric DM + stochastic interstellar DM}
    \includegraphics[width=\linewidth]{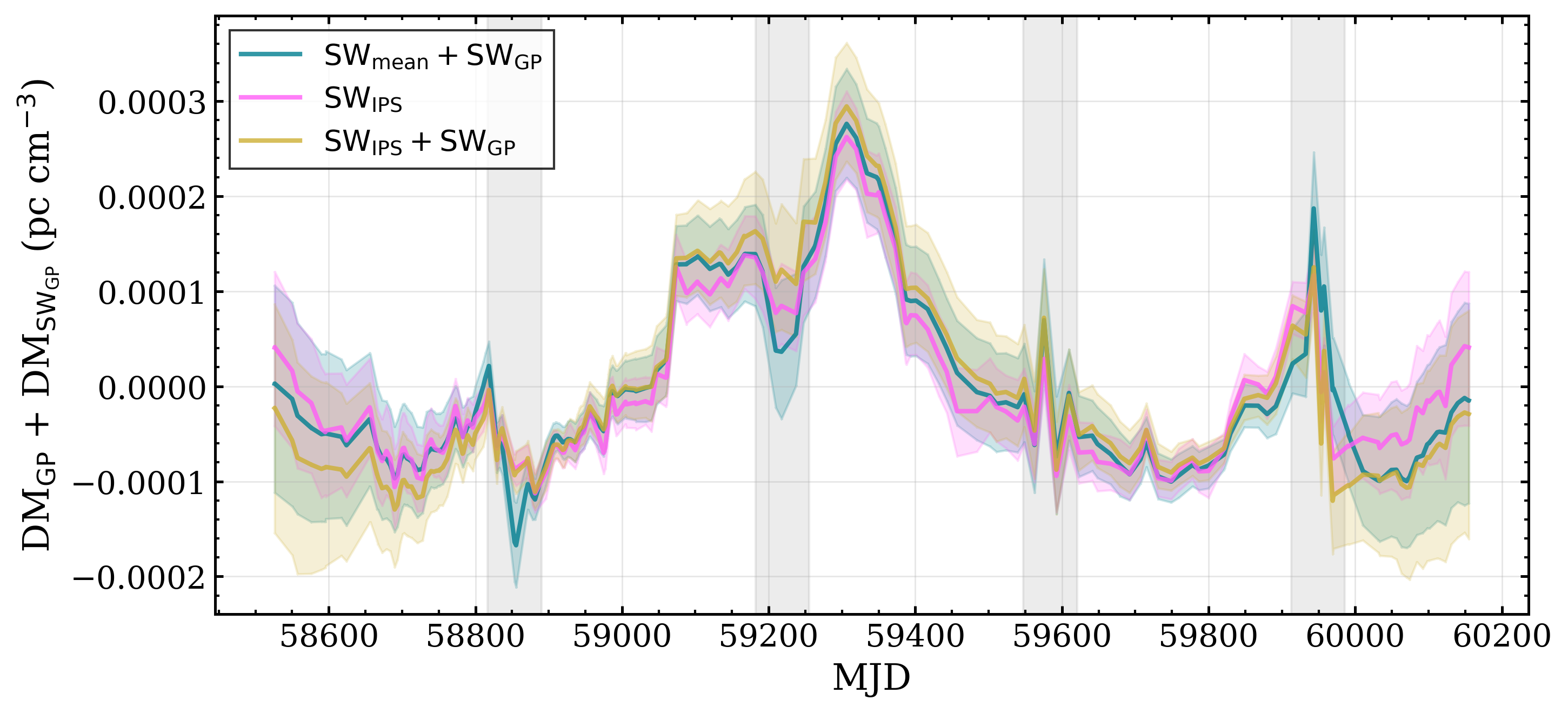}
    
  \end{subfigure}
  \begin{subfigure}{0.49\textwidth}
    \caption{Stochastic common red noise time delay}
    \includegraphics[width=\linewidth]{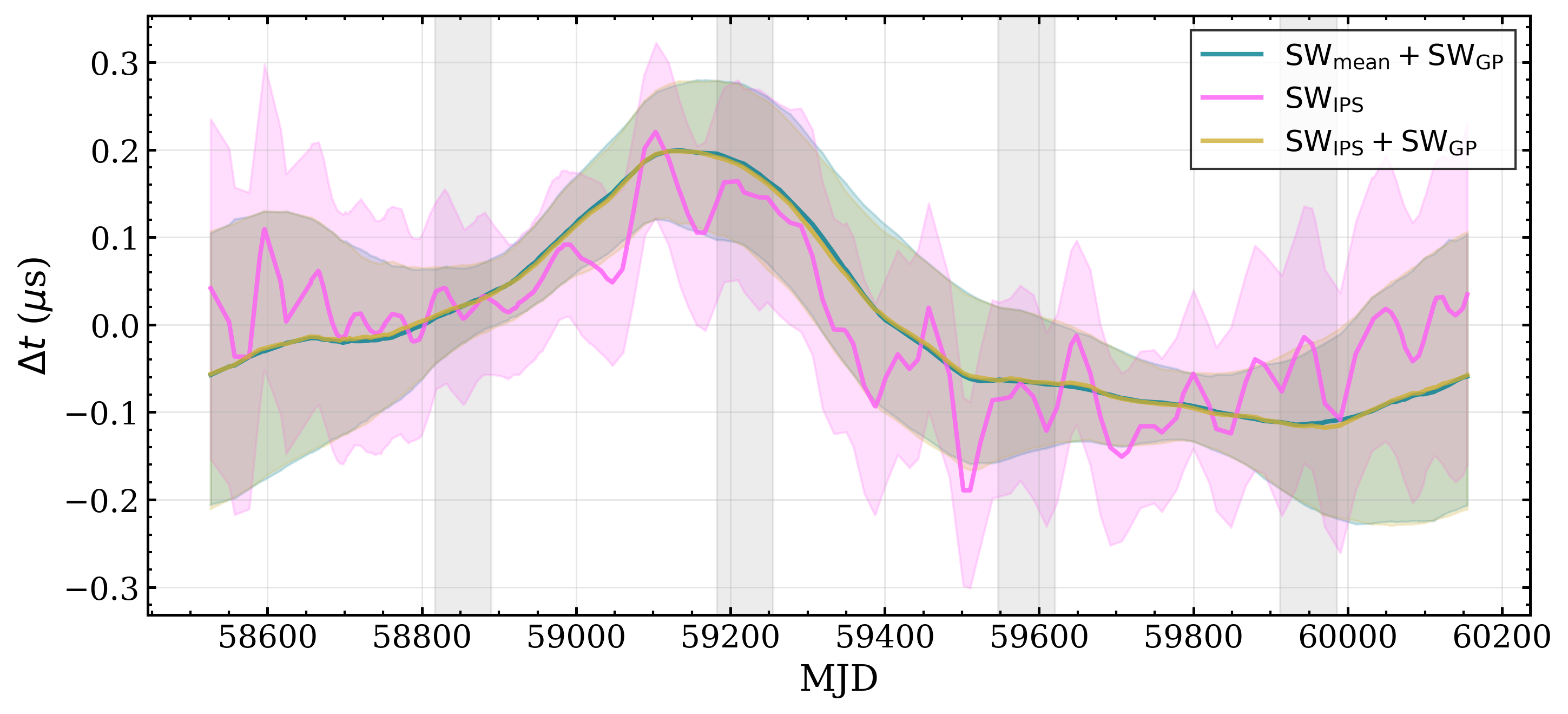}
    
  \end{subfigure}

  \caption{Gaussian-process time-domain realisations for various noise processes for the PSR\,J1909$-$3744. Top left: the total heliospheric DM - DM$_{SW}$ (i.e. DM$_{\mathrm{SW_{mean}}}$ or DM$_{\mathrm{SW_{IPS}}}$ + DM$_{\mathrm{SW_{GP}}}$); top right: the inferred interstellar DM variations / DM$_{\mathrm{GP}}$; bottom left: then net DM variations with contributions from the interstellar (DM$_{\mathrm{GP}}$)  and stochastic heliospheric DM variations (DM$_{\mathrm{SW_{GP}}}$); bottom right: the time delays of a common process, reflective of potential GW signals. The teal, pink and gold curves show results when using the ``SW$_{\mathrm{mean}}$+SW$_{\mathrm{GP}}$", ``SW$_{\mathrm{IPS}}$" and ``SW$_{\mathrm{IPS}}$+SW$_{\mathrm{GP}}$" models for heliospheric delays, respectively. The corresponding $1\sigma$ error regions are shown as filled areas in corresponding colours. The grey regions representing the epoch when the PSR~J1909$-$3744 is within 40$^{\circ}$ of the Sun. For interstellar DM variations, a clear change in DM can be observed at small solar elongations (shaded grey bands).}
  \label{fig:time_real}
\end{figure*}

\section{Results}
\label{results}
We present the results of directly using IPS-UCSD derived 3D heliospheric delays in the search for a CRN on MPTA DR2 and examine its impact on various noise parameters and recovered spatial correlations. Building on the noise processes established in MPTA DR2, we perform the uncorrelated CRN search using 3 different heliosphere models: a) ``SW$_{\mathrm{mean}}$+SW$_{\mathrm{GP}}$'', b) ``SW$_{\mathrm{IPS}}$'', and c) ``SW$_{\mathrm{IPS}}$+SW$_{\mathrm{GP}}$''.

For the ``SW$_{\mathrm{IPS}}$'' model, we first removed the outliers from the $\mathrm{DM_{IPS}}$. Fig. \ref{fig:interpolation} shows this removal of outliers for PSR~J1909$-$3744. Applying this technique to all 83 MSPs in MPTA DR2 typically yields 2-4 outliers per pulsar. We then added excess DM due to heliospheric delays in the pulsar-Earth beyond 3\,AU using the correction procedure described in Section \ref{IPS-SWM} to get $\mathrm{DM_{IPS,final}}$.


\subsection{Parameter estimation and time-domain realisation comparison of various noise processes}

Fig. \ref{fig:noise_param} shows the posterior probability distributions for the parameters describing the stochastic interstellar DM variations  and the heliospheric density at 1\,AU for PSR~J1909$-$3744 and the uncorrelated CRN when using various heliosphere models. The power-law PSD describing the stochastic interstellar DM variations is shallower for PSR~J1909$-$3744 when using the ``SW$_{\mathrm{IPS}}$'' model, whereas for ``SW$_{\mathrm{mean}}$+SW$_{\mathrm{GP}}$'' and ``SW$_{\mathrm{IPS}}$+SW$_{\mathrm{GP}}$'' cases, the DM variations are steeper and nearly identical and consistent with the expectation of Kolmogorov turbulence. Similarly, the inferred properties of the power-law describing the heliospheric density at 1\,AU for the ``SW$_{\mathrm{IPS}}$+SW$_{\mathrm{GP}}$'' case shows a broad distribution centred around a spectral exponent of zero, and the amplitude is smaller than when using ``SW$_{\mathrm{mean}}$+SW$_{\mathrm{GP}}$''. In the latter case, the stochastic heliospheric density variation is well constrained, supporting a loud and steeper noise. The uncorrelated CRN has a very shallow spectral index and high amplitude when using ``SW$_{\mathrm{IPS}}$'', whereas for the ``SW$_{\mathrm{mean}}$+SW$_{\mathrm{GP}}$'' and ``SW$_{\mathrm{IPS}}$+SW$_{\mathrm{GP}}$'' cases, the recovered noise is steeper and nearly identical. The CRN spectral index is consistent with the expectation for a GWB from SMBHBs for the noise models that include SW$_{\mathrm{GP}}$.

Fig. \ref{fig:time_real} shows the time-domain realisation of various noise processes for PSR~J1909$-$3744. On average, delays from different heliosphere models are similar. Because PSR~J1909$-$3744 lies below the solar pole, the solar activity cycle is more evident: heliospheric density above/below the pole of the Sun varies more over the solar cycle, till ELAT < 25$^{\circ}$ \citep{Tiburzi_2021, Susarla_2024, waszewski2025latitudinaldependencesolarwind}. The heliospheric density on the equatorial plane of the Sun remains relatively stable and this can be seen in the MPTA DR2. Two pulsars PSR~J1022$+$1001 (ELAT = -0.06$^{\circ}$) and PSR~J1730$-$2304 (ELAT = 0.189$^{\circ}$) are very close to the ecliptic plane and has low heliospheric variability, i.e., no evidence of SW$\mathrm{_{GP}}$.

The reconstructed DM variations are smoother when using ``SW$_{\mathrm{mean}}$+SW$_{\mathrm{GP}}$'' and ``SW$_{\mathrm{IPS}}$+SW$_{\mathrm{GP}}$'' than when using ``SW$_{\mathrm{IPS}}$'' alone. With ``SW$_{\mathrm{IPS}}$'' the stochastic interstellar DM variations show pronounced dips and rises near MJD $\sim$ 59210 and 59950, which correspond to epochs when the pulsar is very close to the Sun and heliospheric delays are expected to be larger. When examining the sum of the stochastic variations in the interstellar and the heliospheric DM for the different heliosphere$-$model cases, the rises and dips in the above summed DM variations occur at the same MJDs (close approaches to the Sun), indicating that the unmodelled heliospheric delays are leaking into the stochastic interstellar DM variations when using ``SW$_{\mathrm{IPS}}$'' alone. 

The reconstructed uncorrelated CRN shown in Fig. \ref{fig:time_real} is smoother and very similar for ``SW$_{\mathrm{mean}}$+SW$_{\mathrm{GP}}$'' and ``SW$_{\mathrm{IPS}}$+SW$_{\mathrm{GP}}$'', but appears more rapidly varying when using only ``SW$_{\mathrm{IPS}}$''. This faster variation in the recovered CRN is consistent with residual high-frequency power left by subtracting the rapidly varying ``SW$_{\mathrm{IPS}}$'' directly from the residuals: because ``SW$_{\mathrm{IPS}}$'' contains substantial high-frequency variations, the CRN becomes shallower (as we show in Fig. \ref{fig:noise_param}) to accommodate the non-negligible power at frequencies up to $\sim $1\,yr$^{-1}$.

We inspected the effect of using different heliospheric models on various noise process parameters. We found that for some pulsars close to the ecliptic plane, noise processes are changing when using the ``SW$\mathrm{_{GP}}$'' model, because of the same high-frequency structure that we identified for PSR~J1909$-$3744 and in the recovered common-spectrum process. For PSR~J1811$-$2405 (ELAT = -0.67$^{\circ}$), PSR~J1614$-$2230 (ELAT = -1.26$^{\circ}$), PSR~J1643$-$1224 (ELAT = 9.78$^{\circ}$), and PSR~J1545$-$4550 (ELAT = -25.29$^{\circ}$), DM$_{\mathrm{GP}}$ becomes shallower when using the IPS model, similar to the PSR~J1909$-$3744. For PSR~J1811$-$2405 (ELAT = -0.67$^{\circ}$), PSR~J1327$-$0755 (ELAT = 1.2$^{\circ}$), and PSR~J1614$-$2230 (ELAT = -1.26$^{\circ}$), the SW$_{\mathrm{GP}}$ also changes.   
For the remaining pulsars, the various noise parameters posteriors (including white noise parameters) overlap with those of the ` `SW$\mathrm{_{mean}}$ + SW$\mathrm{_{GP}}$'' model. Although the ``SW$\mathrm{_{IPS}}$'' model gives the same result in these pulsars, our common noise analysis shows that the errors can appear as a common process. For this reason, we would not recommend using this ``SW$\mathrm{_{IPS}}$'' model as low-amplitude errors might not affect single pulsar analyses, but appear in the more sensitive common-noise search. 

\subsection{Spatial correlations}
\subsubsection{MPTA DR2}
\begin{figure}
  \centering

  \begin{subfigure}[b]{\linewidth}
    \includegraphics[width=\linewidth,keepaspectratio]{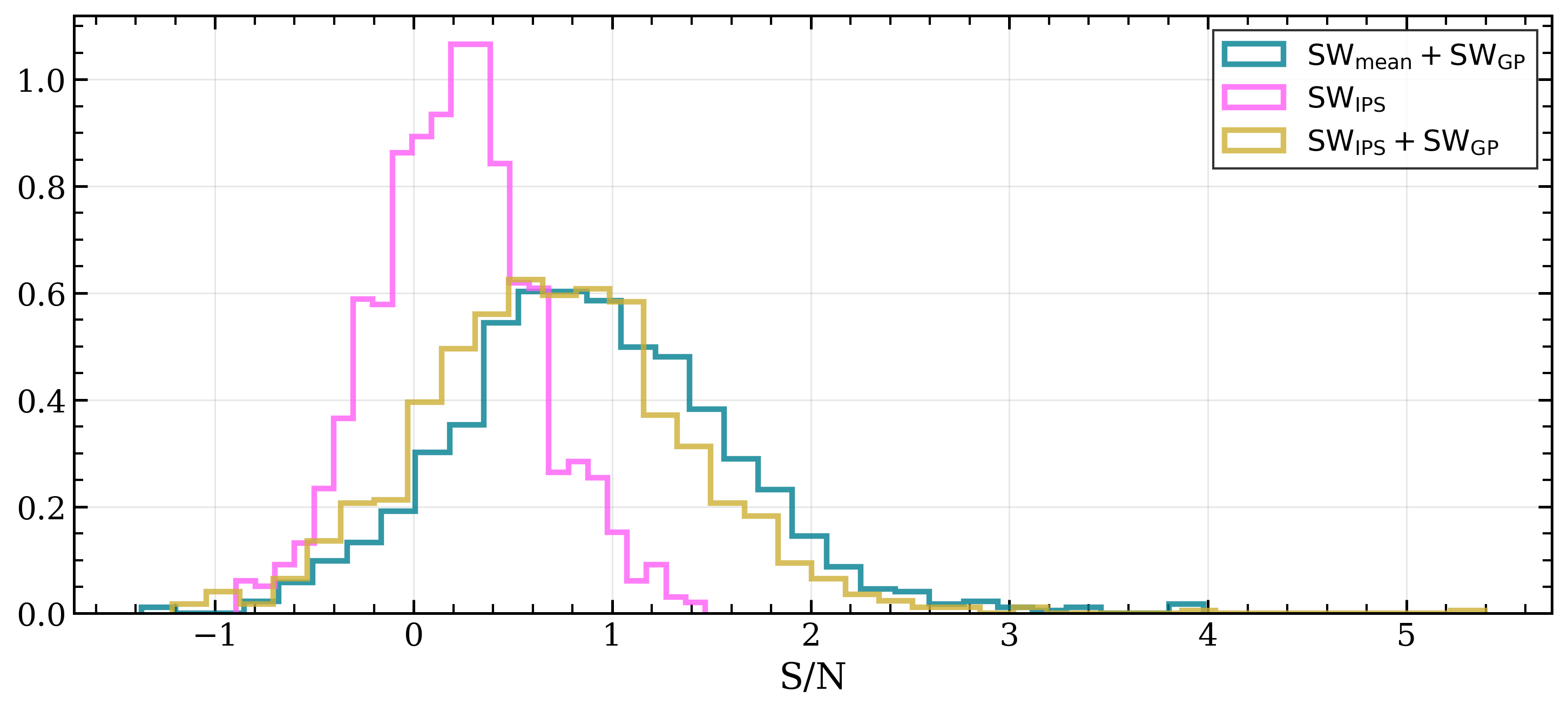}
    \caption{Hellings-Downs correlations} \label{fig:HD}
  \end{subfigure}

  \begin{subfigure}[b]{\linewidth}
    \includegraphics[width=\linewidth,keepaspectratio]{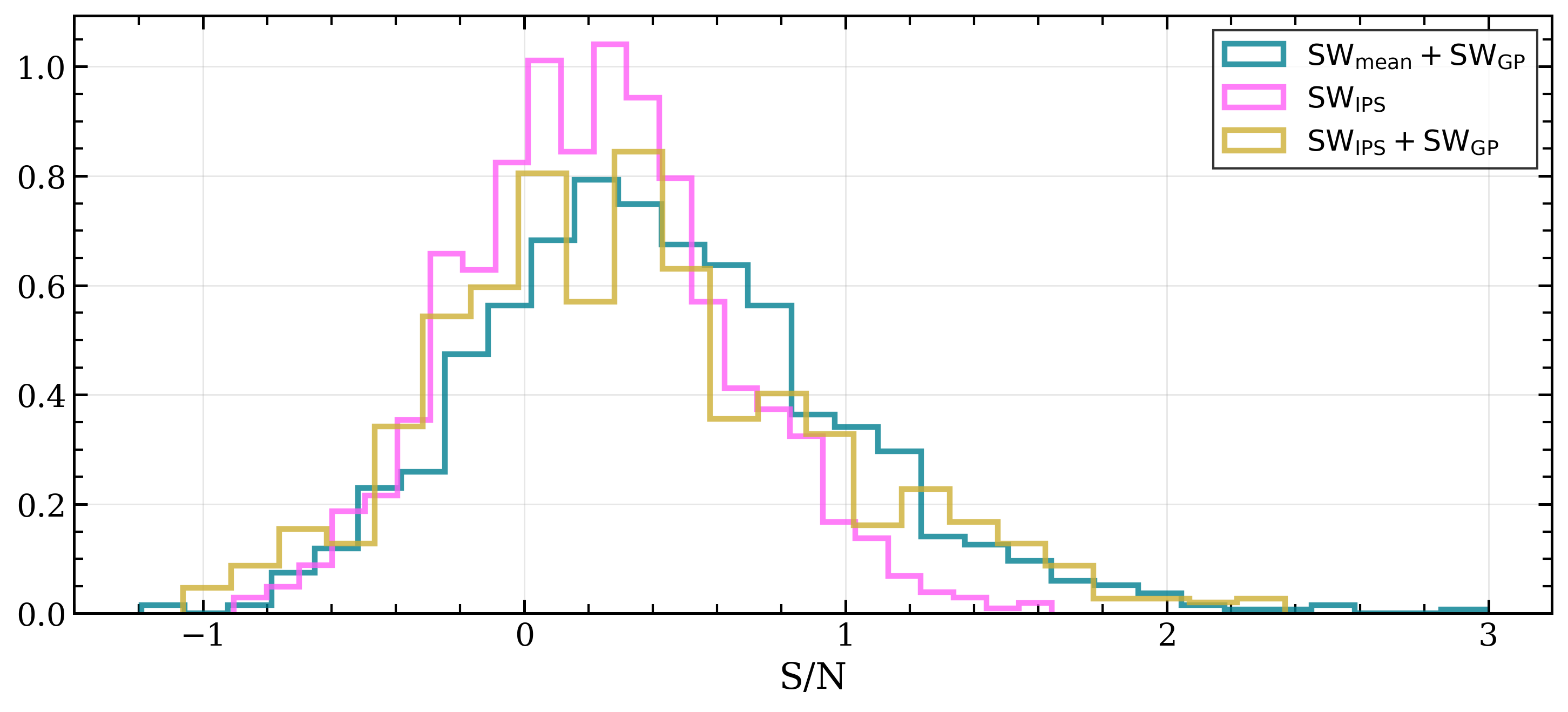}
    \caption{Dipole correlations} \label{fig:dipole}
  \end{subfigure}

  \begin{subfigure}[b]{\linewidth}
    \includegraphics[width=\linewidth,keepaspectratio]{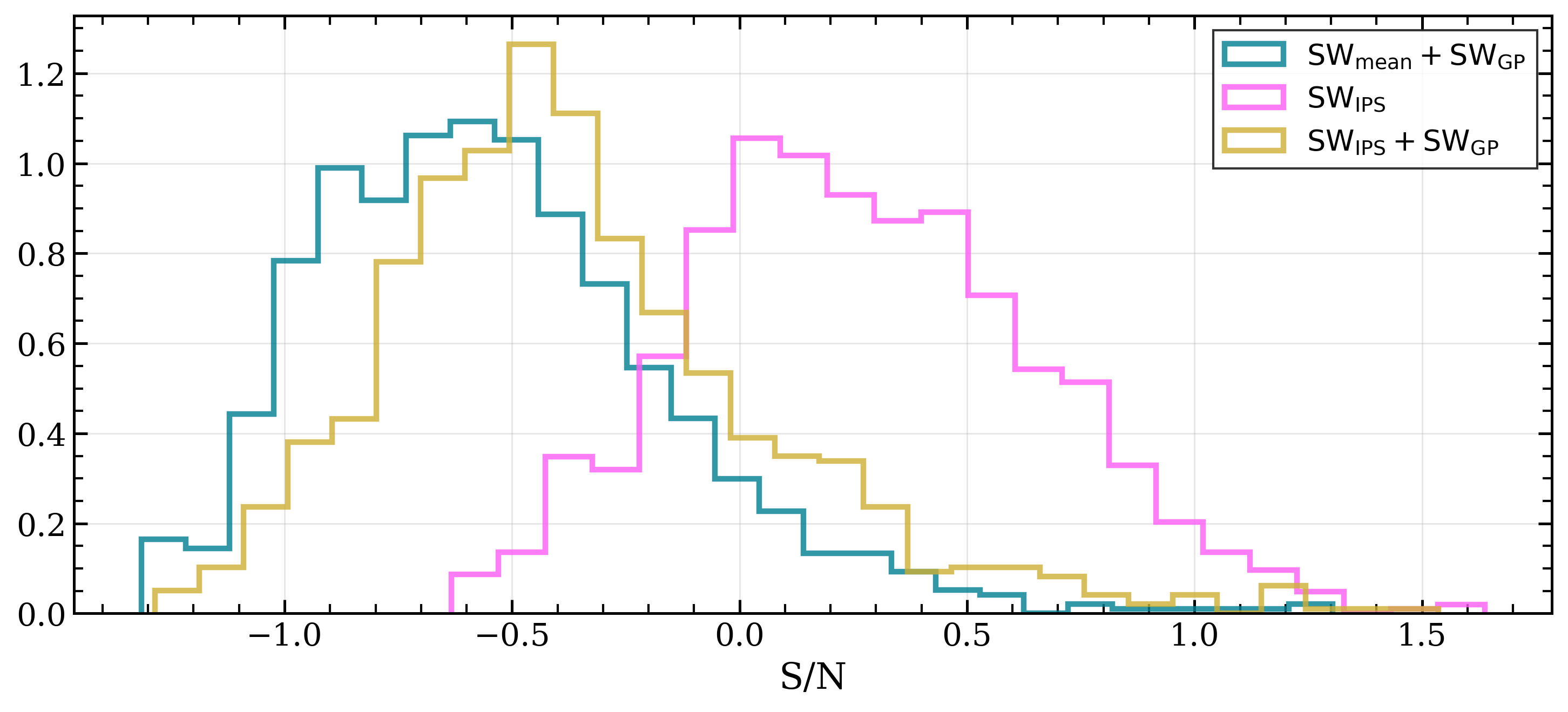}
    \caption{Monopole correlations} \label{fig:monopole}
  \end{subfigure}

  \caption{Histogram of the strength of spatial correlations assuming Hellings-Downs, dipole and monopole ORFs for the MPTA DR2, computed while utilising different heliosphere models. The S/N values indicate the signal-to-noise ratio strength of the recovered correlations for each heliosphere model. The teal, pink and gold curves show results for the ``SW$_{\mathrm{mean}}$+SW$_{\mathrm{GP}}$'', ``SW$_{\mathrm{IPS}}$'' and ``SW$_{\mathrm{IPS}}$+SW$_{\mathrm{GP}}$'' heliosphere model, respectively.}
  \label{fig:cor_real}
\end{figure}

Fig.\ref{fig:cor_real} shows histograms of the measured correlation strengths obtained using different heliosphere models, computed with the OS. For each trial we draw a random sample from the posterior distribution and use the corresponding noise-parameter values, following a similar approach to \citet{Vigeland_2018, Miles_2024} . For each parameter combination we compute the OS signal-to-noise ratio (S/N) assuming ORFs corresponding to HD, dipole and monopole correlations.

The HD distribution for ``SW$_{\mathrm{mean}}$+SW$_{\mathrm{GP}}$'' exhibits a tail extending to $\sim$3\,$\sigma$ as shown in Fig. \ref{fig:HD}, consistent with \citet{Miles_2024}. The HD S/N is substantially reduced when using ``SW$_{\mathrm{IPS}}$'' alone, whereas ``SW$_{\mathrm{IPS}}$+SW$_{\mathrm{GP}}$'' yields the S/N distribution with a spread similar to ``SW$_{\mathrm{mean}}$+SW$_{\mathrm{GP}}$''. This feature is attributable to the recovered uncorrelated CRN having a very shallow spectrum when ``SW$_{\mathrm{IPS}}$'' is used without the GP component.

Fig. \ref{fig:dipole} shows the dipole S/N distributions which are similar across all three heliosphere models, with ``SW$_{\mathrm{mean}}$+SW$_{\mathrm{GP}}$'' and ``SW$_{\mathrm{IPS}}$+SW$_{\mathrm{GP}}$'' showing slightly broader distributions.

The monopole results are noteworthy as plotted in Fig. \ref{fig:monopole}: ``SW$_{\mathrm{IPS}}$'' produces more positive monopole S/N values compared with ``SW$_{\mathrm{mean}}$+SW$_{\mathrm{GP}}$'' and ``SW$_{\mathrm{IPS}}$+SW$_{\mathrm{GP}}$'', which tend toward negative (anti-correlated) values. A negative monopole S/N indicates that a larger fraction of pulsar pairs exhibit anti-correlation.

\subsubsection{Simulated Data}

\begin{figure}
  \centering

  \begin{subfigure}[b]{\linewidth}
    \includegraphics[width=\linewidth,keepaspectratio]{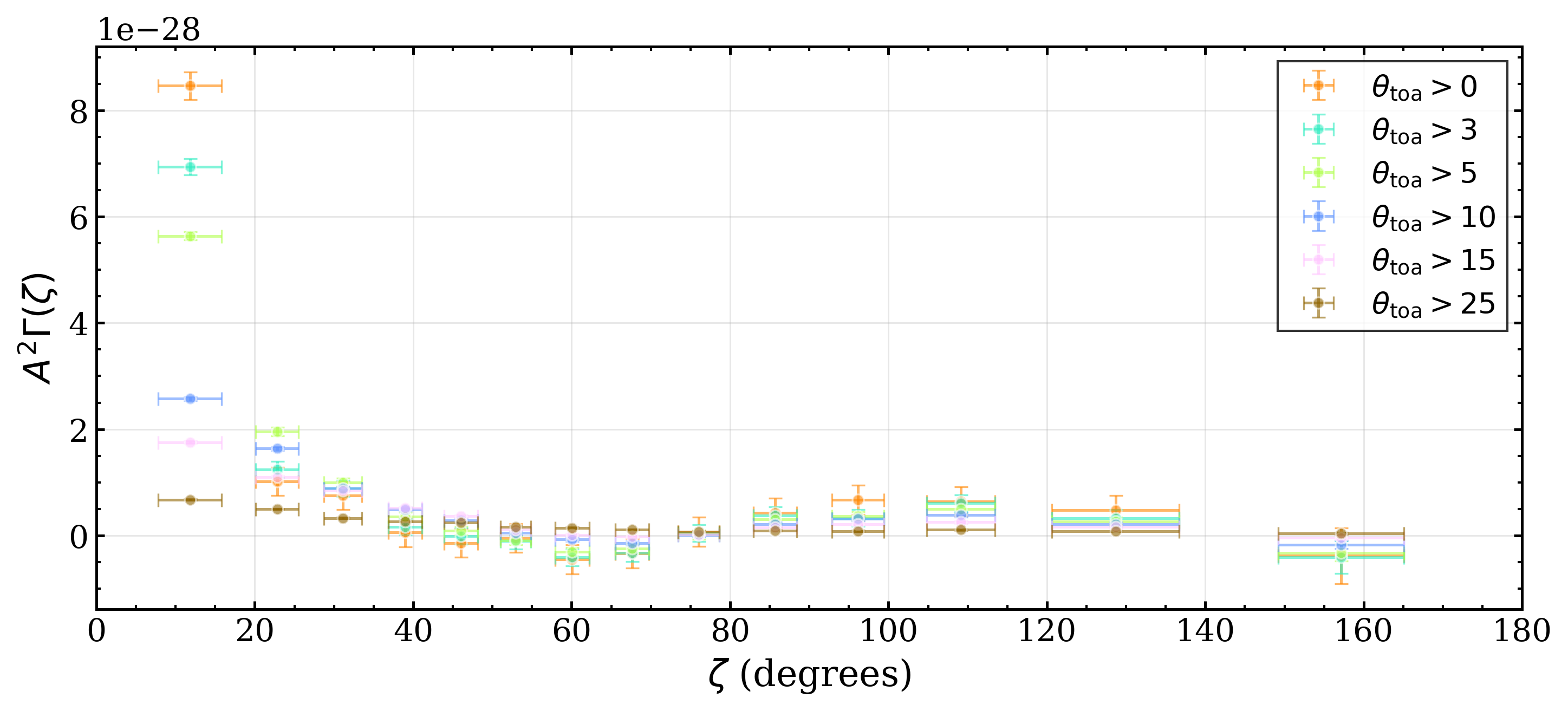}
    \caption{Injected: IPS, Modelled: CRN} \label{fig:SW-corr}
  \end{subfigure}

  \begin{subfigure}[b]{\linewidth}
    \includegraphics[width=\linewidth,keepaspectratio]{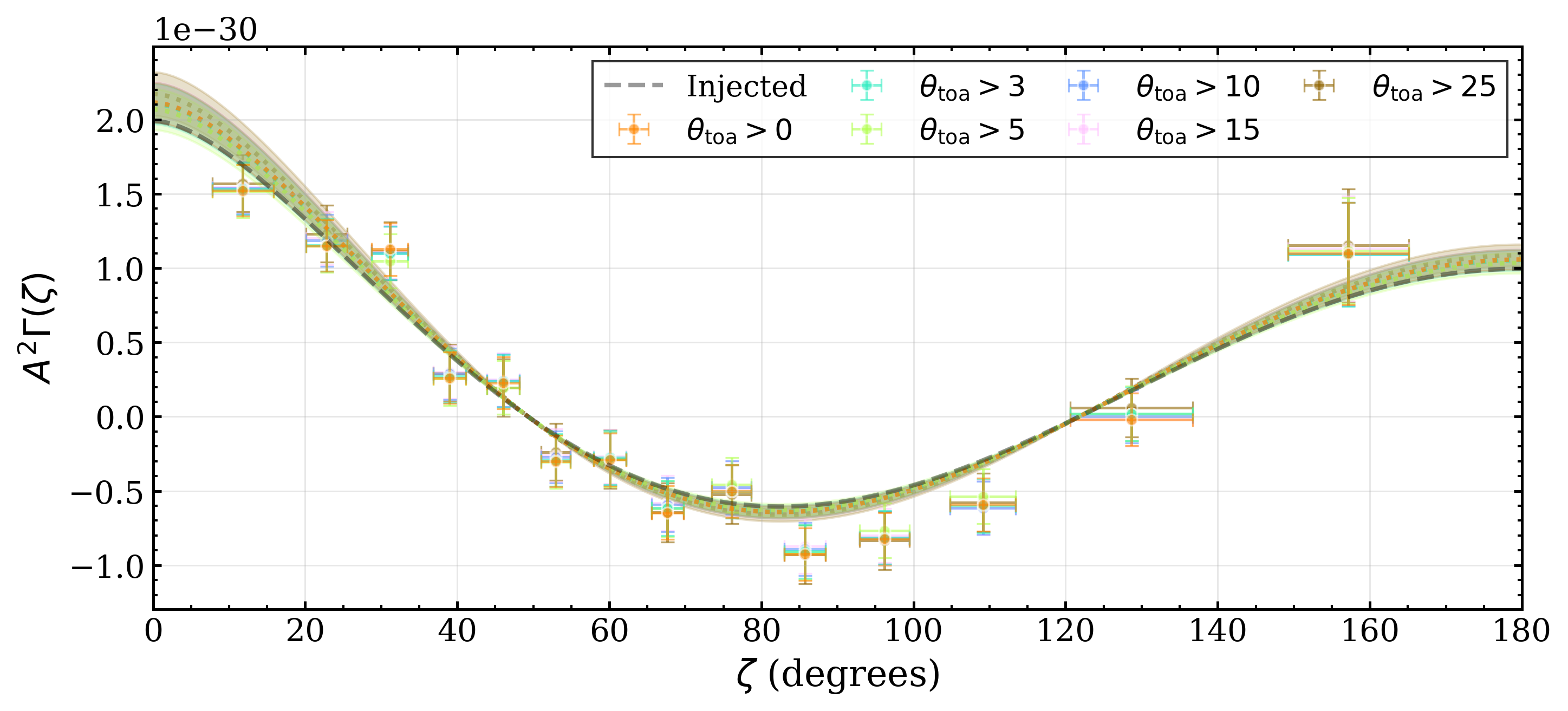}
    \caption{Injected: DM$_{\mathrm{GP}}$+IPS+GWB, Modelled:DM$_{\mathrm{GP}}$+SW$_{\mathrm{mean}}$+SW$_{\mathrm{GP}}$+CRN} \label{fig:SW+GWB-corr}
  \end{subfigure}

  \caption{Measured inter-pulsar spatial correlations induced by the inferred common red noise component in the simulated PTA mock dataset. Top: inter-pulsar correlation obtained using \textbf{simulation 1} where in the simulated PTA only heliospheric delays present (based on IPS-UCSD 3D tomography). These delays are modelled as a uncorrelated common red noise. Bottom: inter-pulsar correlation obtained using \textbf{simulation 2} where in the simulated PTA along with heliospheric delays (IPS-UCSD 3D tomography as a reality), stochastic interstellar DM variations and GWB are also present. The GWB is modelled as a uncorrelated common red noise. Different colours represent the cases where the uncorrelated common red noise has been searched while ignoring the ToAs within certain solar elongation ($\theta_{toa}$).}
  \label{fig:sim_corr}
\end{figure}

\begin{figure*}
    \centering
     \includegraphics[width=\linewidth]{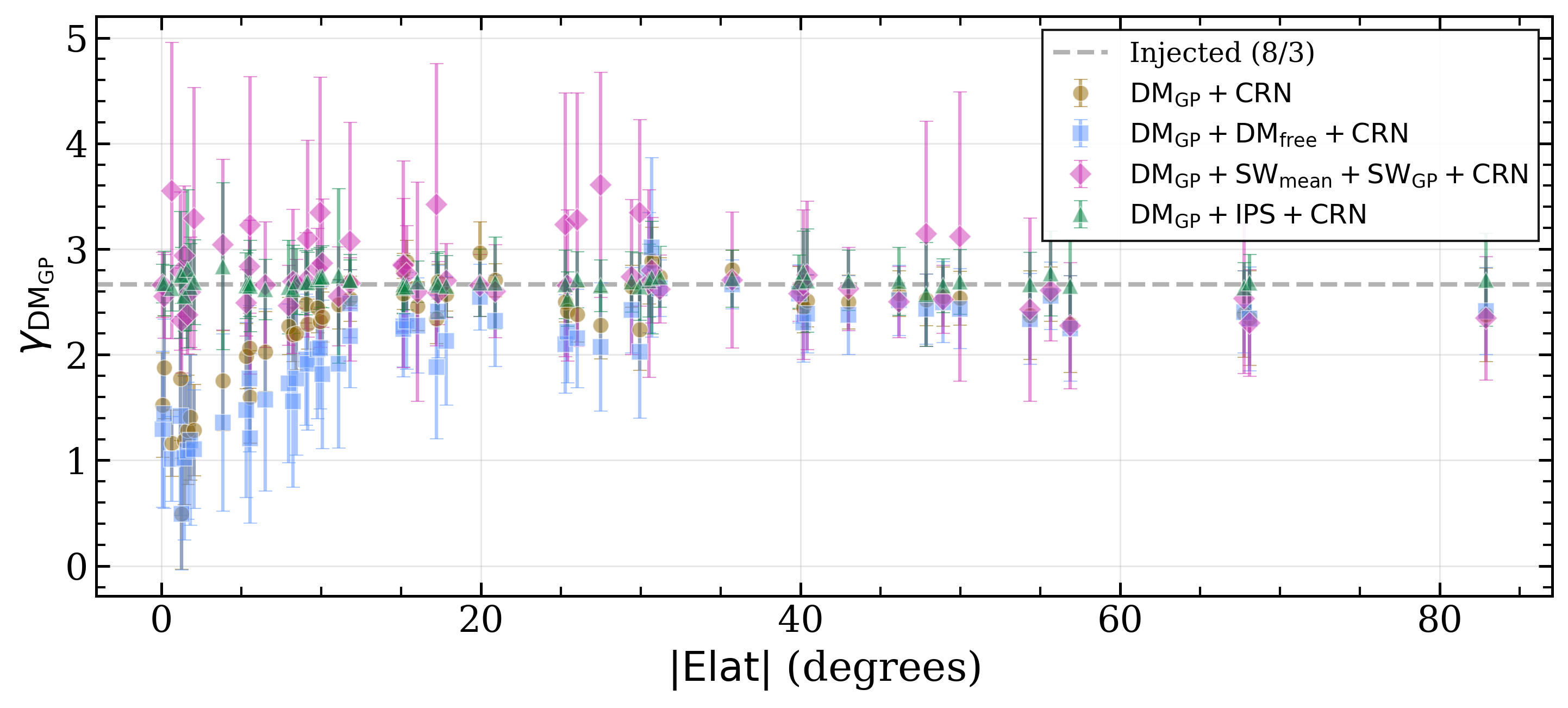}
    \caption{Recovered spectral index of the stochastic interstellar DM  variations for various pulsars as a function of ecliptic latitude (ELAT), where the injected value is 8/3 (Kolmogorov turbulence) using the \textbf{simulation 2} mock PTA dataset. Different colours show the results from different noise models. For all these cases, the injected signals for the simulation were the stochastic interstellar DM variations, white noise, GWB, and IPS-UCSD 3D reconstructed heliosphere model.}
    \label{fig:simdm}
\end{figure*}
 
\begin{figure*}
    \centering
     \includegraphics[width=\linewidth]{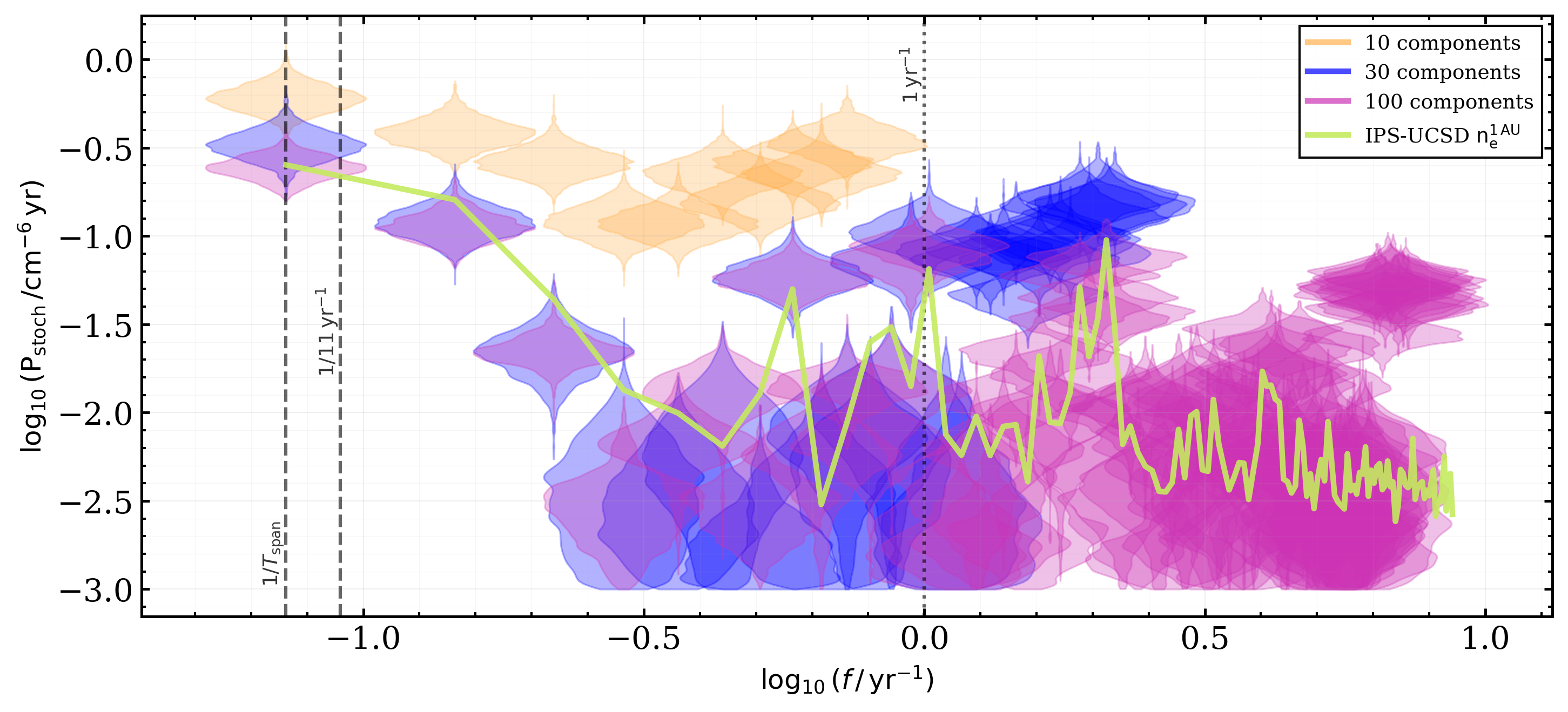}
    \caption{Recovered common PSD of the stochastic heliospheric density variations ($\mathrm{n_{e}^{1\,AU}}$) assuming a free spectrum rather than a power law from \textbf{simulation 3} mock PTA dataset. The simulated dataset contains heliospehric delays based on the IPS-UCSD 3D tomography heliosphere model but modelled using the ``SW$_{\mathrm{mean}}$+SW$_{\mathrm{GP}}$'' model. The y-axis shows the power in the log-scale whereas the x axis shows the Fourier components of the stochastic variation in the log-scale. The green curve shows the injected $\mathrm{n_{e}^{1\,AU}}$. Orange, blue and magenta represent the cases when using various Fourier components to model the stochastic heliospheric density variations, which corresponds to 10, 30, and 100 components respectively. The results demonstrate clear deviations from a simple power-law model, evident in the variable distribution of power across frequency bins. The recovered power depends on the number of frequency components modelled, particularly when frequencies $>1$\,yr$^{-1}$ are not modelled.}
    \label{fig:free-spec}
\end{figure*}

We conducted analyses of simulated MPTA-like mock datasets to assess the impact of the current PTA spherically symmetric time varying heliosphere model in ideal scenarios (while using IPS-UCSD 3D reconstructed heliosphere). The results of the simulated datasets described in Section \ref{simulation-section} are summarised below.

\begin{enumerate}[leftmargin=*]
\item \textbf{Simulation 1:} This dataset contains IPS and white noise - efac. We model uncorrelated CRN and fixing white noise to injected value, i.e., EFAC = 1,  to avoid any leakage of misspecification of heliosphere delay in white noise parameter. We, then utilise the OS code to measure the inter-pulsar spatial correlations. Separate CRN searches were performed on subsets of the simulation, with each subset formed by removing ToAs with solar elongation $\theta_{\rm toa}$ (separation angle from the Sun) within cutoffs $\{0^\circ,\,3^\circ,\,5^\circ,\,10^\circ,\,15^\circ,\,25^\circ\}$. In all subsets, the recovered uncorrelated CRN had a shallow spectral index (ranging $\gamma_{\rm CRN}\sim0.6$--$1.8$). We adopted the mean spectral index and re-ran the CRN search with a fixed $\gamma_{\rm CRN}=1.35$ to ensure a consistent amplitude and raw spatial correlation comparison. After this noise analysis, the OS was used to compute inter-pulsar correlations and to normalise the autocovariances to be at the same as the measured common-red-noise amplitude level. shown in Fig.\ref{fig:SW-corr}. As the masking cutoff increases, the heliosphere-induced correlation decreases and the overall correlation moves towards zero; the correlation is dominated by pulsars that approach close to the Sun.

\item \textbf{Simulation 2:} This dataset contains IPS, DM$_{\mathrm{GP}}$, a GWB, and white noise - efac. The interstellar DM variations were modelled with DM$_{\mathrm{GP}}$ model together with CRN, while the heliosphere was intentionally misspecified in three ways: (i) no heliosphere model, (ii) ``SW$_{\mathrm{mean}}$+SW$_{\mathrm{GP}}$'', and (iii) DM$_{\mathrm{free}}$ (see Section~\ref{simulation-section}), along with the ideal case ``SW$_{\mathrm{IPS}}$'' where we have subtracted the injected IPS. 
Fig. \ref{fig:SW+GWB-corr} shows that when the heliospheric delays are modelled using the ``SW$_{\mathrm{mean}}$+SW$_{\mathrm{GP}}$'' model, the recovered GWB correlation matches the injected signal even without masking ToAs within given solar-elongation cutoffs. Fig. \ref{fig:simdm} presents recovered spectral indices ($\gamma_{\mathrm{DM_{GP}}}$) of the power-law describing the stochastic interstellar DM variations: with a badly specified heliosphere model, the recovered $\gamma_{\mathrm{DM_{GP}}}$ shows an ecliptic-angle dependence (pulsars near the Sun display artificially shallow spectra), whereas the ``SW$_{\mathrm{mean}}$+SW$_{\mathrm{GP}}$'' model recovers the injected interstellar DM variation spectral index. Across all pulsars, the recovered white noise parameter - EFAC remain within 1$\sigma$ of the injected values, for various misspecifications of the heliospheric delays.

\item \textbf{Simulation 3:} This dataset spans $\sim$13-years and contains IPS-UCSD 3D reconsturcted heliospheric delays and white noise - EFAC. The heliospheric delays are modelled as a common noise process using ``SW$_{\mathrm{mean}}$+SW$_{\mathrm{GP}}$'' with a free-spectrum parameterisation (power fitted independently in each frequency bin) PSD along with white noise - EFAC and ECORR. The ECORR has been modelled to account for the fact that the free spectrum model does not go to infinite Fourier frequencies but the IPS-UCSD injected heliospheric delay does, effectively, so we needed a model for the high-Fourier frequencies. We compare models with 10, 30 and 100 Fourier frequency components. Fig. \ref{fig:free-spec} displays the recovered free spectrum of $\mathrm{n_{e}^{1\,AU}}$: the spectrum changes notably at low Fourier frequencies between the different models, and the shape of the spectrum depends on the number of components and on whether bins sample frequencies above or below $1\,\mathrm{yr}^{-1}$ (further discussed in Section \ref{discussion}). Using Eq. \ref{DM_sph} to compute $\mathrm{n_{e}^{1\,AU}}$ from the 13-year DM$_{IPS,final}$ series and taking its power spectrum, the recovered spectrum agrees with the free-spectrum fit for frequencies > $1\,\mathrm{yr}^{-1}$.  The white-noise parameters are well recovered using a sufficiently high number of frequency components. We show that this number is approximately 100 for our simulated 13-year data set. When too few components are used, the white-noise parameters for low ecliptic latitude pulsars ($|ELAT| < 6^{\circ}$) are higher than injected.

\end{enumerate}

\section{Discussion and Summary}
\label{discussion}

We tested whether the IPS-based heliosphere reconstruction can be directly used to model the variable heliosphere in PTA datasets. We also tested the performance of current PTA noise models under simulated datasets where the IPS-UCSD 3D heliosphere reconstruction was injected as a realistic representation of the heliosphere. We found that for the MPTA DR2 dataset:

\begin{enumerate}[leftmargin=*]
    \item The IPS-UCSD based 3D heliospheric reconstruction does not provide an improved model for heliospheric delays in PTA analyses.  The recovered noise parameters are altered by the IPS-derived heliosphere, including (a) divergence of the spectral slope of the interstellar DM variations away from the expected $8/3$ value (see Figs. \ref{fig:noise_param}), and (b) a much shallower spectral slope for the common red noise as shown in Figs. \ref{fig:noise_param}, \ref{fig:time_real}. The change in noise model parameters are more pronounced for some MPTA pulsars within $|ELAT| < 16^{\circ}$, which demonstrates that these pulsars show strong sensitivity to heliospheric delays.  
    \item Related to point (b) above, because the IPS-UCSD-based 3D heliosphere reconstructions alter the spectral properties of the recovered common red noise, they also reduce the recovered S/N of HD correlations. This reduction arises because of the additional short timescale structure induced in the common red noise which originates from misspecification of heliospheric delays, and is unrelated to the sought-after GW signal (see Fig. \ref{fig:cor_real}).
    \item In the absence of any other noise source in the PTA dataset, misspecification of heliospheric delays can affect the recovered spatial correlations. This was demonstrated by measuring the spatial correlation induced by the IPS-UCSD 3D reconstructed heliosphere, shown in Fig. \ref{fig:SW-corr}. However, the correlation changes again when other noise sources (specifically a DM Gaussian process) are present and modelled in the PTA, and we can recover the injected GWB correlation as shown in Fig. \ref{fig:SW+GWB-corr}.
    \item Assuming the MPTA noise models, removing ToAs within certain separation angles from the Sun does not significantly alter the recovered correlations. This suggests that for MPTA-like datasets, excising ToAs recorded at low solar elongation is not an effective strategy to mitigate heliospheric delays.
    \item Simulations show that misspecification of the heliosphere model in the presence of other noise processes in PTA datasets biases the parameters of these noise models for low ecliptic latitude pulsars, specifically the interstellar DM variations, as shown in Fig. \ref{fig:simdm}. However, the recovered spectral properties of the GWB are consistent with the injected values.
    \item Using a time$-$variable spherically symmetric model (``SW$_{\mathrm{mean}}$ + SW$_{\mathrm{GP}}$'') is adequate to model the heliospheric delays in the MPTA dataset along with the GWB and interstellar DM variations.
    \item It is important to model the high-frequency components of heliospheric delays variations when using the Fourier domain kernel, as shown in Fig. \ref{fig:free-spec}. Only modelling low-frequency Fourier components leads to gross mis-estimation of the time-variable heliosphere properties. The recovered white noise parameters also get affected when using low-frequncey Fourier components for low-ecliptic pulsar, $|ELAT| < 6^{\circ}$. Furthermore, assuming the IPS-UCSD derived 3D reconstruction of the heliosphere is reasonably close to reality, the structure of the power spectrum of heliospheric density variations deviates strongly from a power law under the ``SW$_{\mathrm{mean}}$+SW$_{\mathrm{GP}}$'' model. 
\end{enumerate}

We have tested whether the IPS-based derived heliosphere, reconstructed using UCSD 3D tomography, can be used to model the heliospheric delays when searching for GWB in MPTA DR2. IPS-derived heliospheric models are currently highly competitive  due to their ability to construct plausible three-dimensional models of the time-varying heliosphere. However, directly using ``SW$_{\mathrm{IPS}}$'' alters several noise parameters and the uncorrelated CRN: the recovered CRN becomes shallower, indicating leakage from IPS-derived heliospheric delays that add power at higher Fourier frequencies. The time-domain realisations show the same behaviour: interstellar DM variations and CRN exhibit more structure on short timescales when ``SW$_{\mathrm{IPS}}$'' is used, consistent with leakage of high-frequency power into other noise sources. When using ``SW$_{\mathrm{mean}}$+SW$_{\mathrm{GP}}$'' or ``SW$_{\mathrm{IPS}}$+SW$_{\mathrm{GP}}$'' the recovered noise is consistent, suggesting the stochastic heliospheric delays component absorbs the fast IPS variations in later and prevents contamination of other parameters. The fact that we can measure these differences, for example with PSR~J1909$-$3744, indicates that at least some pulsars are sensitive to errors in such heliosphere models. The consequence of this is that that pulsar-based measurements of the heliospheric delays can be used to improve 3D, time-dependent models of the heliosphere. This is particularly pertinent with the advent of low-frequency instruments such as LOFAR, NENUFAR, and the SKA-Low in PTA experiments, which will provide even higher-precision integrated heliospheric density estimates, and the ongoing efforts to improve space weather models for e.g. coronal mass ejection forecasting.

We also found that the optimal statistic amplitudes for various ORFs (HD, dipole, and monopole) can be biased when the heliosphere model is misspecified: leaked IPS power can dominate the recovered signal and reduce genuine GWB-induced correlations. ``SW$_{\mathrm{IPS}}$'' alone does not adequately model the heliosphere because its fast variations are absorbed into the uncorrelated CRN, reducing the HD S/N.

Simulations show that in the absence of advanced noise models, masking ToAs within certain solar-elongation cutoffs reduces the impact of the heliospheric delays, particularly for pulsar pairs with members close to the ecliptic plane. Notably, although a misspecified heliosphere model affects the recovery of other noise parameters, the ``SW$_{\mathrm{mean}}$+SW$_{\mathrm{GP}}$'' model is able to model the IPS-derived heliospheric delays and yields reasonable recovery of other noise parameters {without} the need to excise ToAs. 

We also show that to account for the full heliospheric delays signal one must model a high number of Fourier-frequency components to capture the fast variations. The power in the SW$_{\mathrm{GP}}$ model depends on the number of frequency components used, which may point to a limitation of Fourier-domain analyses that treat each Fourier frequency bin as independent \citep{crisostomi2025diagonalapproximationsimprovedcovariance}. Time-domain models of the heliosphere may be more appropriate \citep{crisostomi2025diagonalapproximationsimprovedcovariance, hazboun2025nanograv125yeardataset} and could be further tested against the IPS-UCSD 3D reconstructed derived heliosphere in future work. 

The current heliosphere model employed by many PTAs (``SW$_{\mathrm{mean}}$+SW$_{\mathrm{GP}}$'') is adequate for GWB searches in MPTA L-band datasets. More work is needed to address the inclusion of low radio frequencies in PTA datasets where heliospheric delays are a dominant feature \citep[e.g. in LOFAR and future SKA-low pulsar timing data][]{shannon2025skaopulsartimingarray}. Our work also shows that pulsar timing is a sensitive probe of the heliospheric structure. Our observations show that current state-of-the-art IPS models are not yet adequate to fully account for heliospheric delays in pulsar timing. We anticipate that future IPS observations with SKA-low, in conjunction with precision low-frequency timing measurements of pulsars, may be used to produce more accurate models of the 3D heliosphere that can be used in future L-band pulsar timing experiments \citep{tiburzi2025exploringgalacticplasmapulsars}. 

\section*{Acknowledgements}

The authors wish to acknowledge Dr. Bernard V. Jackson for providing the IPS-UCSD 3D tomography data.
The authors also thank Dr. Valentina di Marco for her useful comments on this work.
SM, DJR, MB, RMS, and ADK acknowledge support from the ARC Centre of Excellence for Gravitational Wave Discovery (CE170100004 and CE230100016). RMS acknowledges support through the ARC Future Fellowship FT190100155. MTM acknowledges support from the NANOGrav Collaboration’s National Science Foundation grant. 
The MeerKAT telescope is operated by the South African Radio Astronomy Observatory (SARAO), which is a facility of the National Research Foundation, an agency of the Department of Science and Innovation. PTUSE was developed with the support of the Australian SKA Office and Swinburne University of Technology, with financial contributions from the MeerTime collaboration members. This work used the OzSTAR national facility at Swinburne University of Technology. OzSTAR is funded by Swinburne University of Technology and the National Collaborative Research Infrastructure Strategy (NCRIS).
\section*{Data Availability}

All data used in this work are available courtesy of AAO Data Central (\href{https://datacentral.org.au/}{https://datacentral.org.au/}) at \href{https://doi.org/10.57891/j0vh-5g31}{https://doi.org/10.57891/j0vh-5g31}. The data generated through IPS-UCSD 3D heliosphere reconstruction model and simulations will be made available on reasonable request to the corresponding author.



\bibliographystyle{mnras}
\bibliography{bibfile} 







\bsp	
\label{lastpage}
\end{document}